\documentclass[aps,prb,twocolumn,superscriptaddress,print]{revtex4}
\usepackage{amssymb,amsmath}
\usepackage{graphicx}
\usepackage{multirow}
\usepackage{hyperref}
\usepackage[title]{appendix}
\usepackage{mathtools}
\usepackage{amsmath}
\usepackage{verbatim}
\usepackage{gensymb}
\usepackage{subfigure}
\usepackage{bm}
\usepackage{threeparttable}
\usepackage{booktabs}
\usepackage{braket}
\usepackage{xcolor}
\usepackage[normalem]{ulem}

\begin{document}
\title{Electronic structure of 30{\degree} twisted double bilayer graphene}
\author{Guodong Yu}
\affiliation{Key Laboratory of Artificial Micro- and Nano-structures of Ministry of Education and School of Physics and Technology, Wuhan University, Wuhan 430072, China}
\affiliation{Institute for Molecules and Materials, Radboud University, Heijendaalseweg 135, NL-6525 AJ Nijmegen, Netherlands}
\author{Zewen Wu}
\affiliation{Key Laboratory of Artificial Micro- and Nano-structures of Ministry of Education and School of Physics and Technology, Wuhan University, Wuhan 430072, China}
\author{Zhen Zhan}
\affiliation{Key Laboratory of Artificial Micro- and Nano-structures of Ministry of Education and School of Physics and Technology, Wuhan University, Wuhan 430072, China}
\author{Mikhail I. Katsnelson}
\affiliation{Institute for Molecules and Materials, Radboud University, Heijendaalseweg 135, NL-6525 AJ Nijmegen, Netherlands}
\author{Shengjun Yuan}
\email{s.yuan@whu.edu.cn}
\affiliation{Key Laboratory of Artificial Micro- and Nano-structures of Ministry of Education and School of Physics and Technology, Wuhan University, Wuhan 430072, China}
\affiliation{Institute for Molecules and Materials, Radboud University, Heijendaalseweg 135, NL-6525 AJ Nijmegen, Netherlands}

\begin{abstract}
In this paper, the electronic properties of 30{\degree} twisted double bilayer graphene, which loses the translational symmetry due to the incommensurate twist angle, are studied by means of the tight-binding approximation. We demonstrate the interlayer decoupling in the low-energy region from various electronic properties, such as the density of states, effective band structure, optical conductivity and Landau level spectrum. However, at $Q$ points, the interlayer coupling results in the appearance of new Van Hove singularities in the density of states, new peaks in the optical conductivity and importantly the 12-fold-symmetry-like electronic states. The k-space tight-binding method is adopted to explain this phenomenon. The electronic states at $Q$ points show the charge distribution patterns more complex than the 30{\degree} twisted bilayer graphene due to the symmetry decrease. These phenomena appear also in the 30{\degree} twisted interface between graphene monolayer and AB stacked bilayer.       
\end{abstract}
\maketitle
\section{Introduction}
In theory, a twisted bilayer graphene (TBG) can transform from a crystalline (commensurate configuration) to quasi-crystalline (incommensurate configuration) depending on the twist angle\cite{commensuration_condition}. At the large twist angle ($\theta >$15{\degree}), the TBG has electronic
properties very similar to those of two decoupled graphene monolayers.\cite{tBG_15plus0,tBG_15plus1,tBG_15plus2,tBG_15plus3,tBG_15plus4,tBG_15plus5,tBG_15plus6,tBG_15plus7} The Fermi velocity can be reduced by decreasing the twist angle.\cite{vf_theta0,vf_theta1} At the so-called magic angle ($\theta\sim$ 1.1{\degree}), the Fermi velocity becomes zero and the flat bands appear in the vicinity of the Fermi level\cite{TBG_flat_band,TBG_flatband1}. Accordingly, the TBG at the small twist angle as a model system of strongly correlated electrons has drawn much attention due to the novel electronic properties, such as the unconventional superconductivity\cite{BG_superconducting,TBG_supertivity1,TBG_supertivity2} and correlated insulator phases\cite{TBG_insulator_phase}. 

The 30{\degree} TBG, an incommensurate bilayer configuration, has been grown successfully on H-SiC(0001)\cite{science_QC}, Pt(111)\cite{pnas_QC}, Cu-Ni(111)\cite{cm_QC} and Cu\cite{30tBG_onCu_arXiv,30tBG_on_Cu_ACSNano} surfaces. As the first two dimensional quasicrystal based on graphene, 30{\degree} TBG has received increasing attention both experimentally and theoretically\cite{30TBG_grown,30TBG_localization,30TBG_superlubricity,30TBG_Suzuki,arXiv_QC,TBG_2dM,Yu_QC_BG}. A method to grow high-quality 30{\degree} TBG epitaxially on SiC using borazine as a surfactant has also been proposed\cite{30TBG_grown}. The 12-fold rotation symmetry and quasi-periodicity of 30{\degree} TBG have been demonstrated by various measurements, such as the Raman spectroscopy, low-energy electron microscopy/diffraction (LEEM/LEED), transmission electron microscopy (TEM) and scanning tunneling microscopy (STM) measurements.\cite{pnas_QC,science_QC,cm_QC,30TBG_STM,30TBG_grown} A number of Dirac cones, especially the mirror-symmetric ones, have been observed by the angle resolved photoemission spectroscopy (ARPES) measurements\cite{science_QC,pnas_QC}. The quasicrystalline order in 30{\degree} TBG can induce unique localization of electrons without any extrinsic disorders\cite{30TBG_localization,arXiv_QC}. All these peculiar properties, especially the quasi-periodicity, make 30{\degree} TBG much different from graphene monolayer, although it has electronic properties very similar to those of two decoupled graphene monolayers in the vicinity of the Fermi level.\cite{decoupling_acs}

Recently, the twisted double bilayer graphene (TDBG) consisting of two $AB$-stacked bilayers, has received much attention especially on the properties\cite{TDBG_spin_polarized_supercond,TDBG1,TDBG2,TDBG3,TDBG_insulating} associated with the strongly correlated electrons in the electrically tunable flat band\cite{TDBG_flatband0,TDBG_flatband1,TDBG_flatband2,TDBG_topol}, such as the superconductivity, magnetic phase transition and correlated insulating state. Besides, the TDBG under the electric field was found to be a valley Hall insulator.\cite{TDBG_topol} A generic twisted multilayer graphene, $M$ layers on top of $N$ layers with a twist angle, possesses two topologically nontrivial flat bands, which exhibit a Chern-number hierarchy.\cite{TMG_chern} Similar to the 30{\degree} TBG, the 30{\degree} TDBG is expected to possess some striking properties due to the disappearance of the translational symmetry. More importantly, the successful fabrication of 30{\degree} TBG and accurate determination of some key structural parameters (such as twist angle, stacking order and interlayer spacing) in experiment\cite{TMG_growth,TMG_growth_carbon} ensure the realization of 30{\degree} TDBG in the near future. Therefore, in this paper, we study the effect of the 30{\degree} twisted interface between two AB stacked graphene bilayers. The 30{\degree} twisted interface between graphene monolayer and AB stacked bilayer is also discussed.

\begin{figure*}[!htbp]
\centering
\includegraphics[width=16 cm]{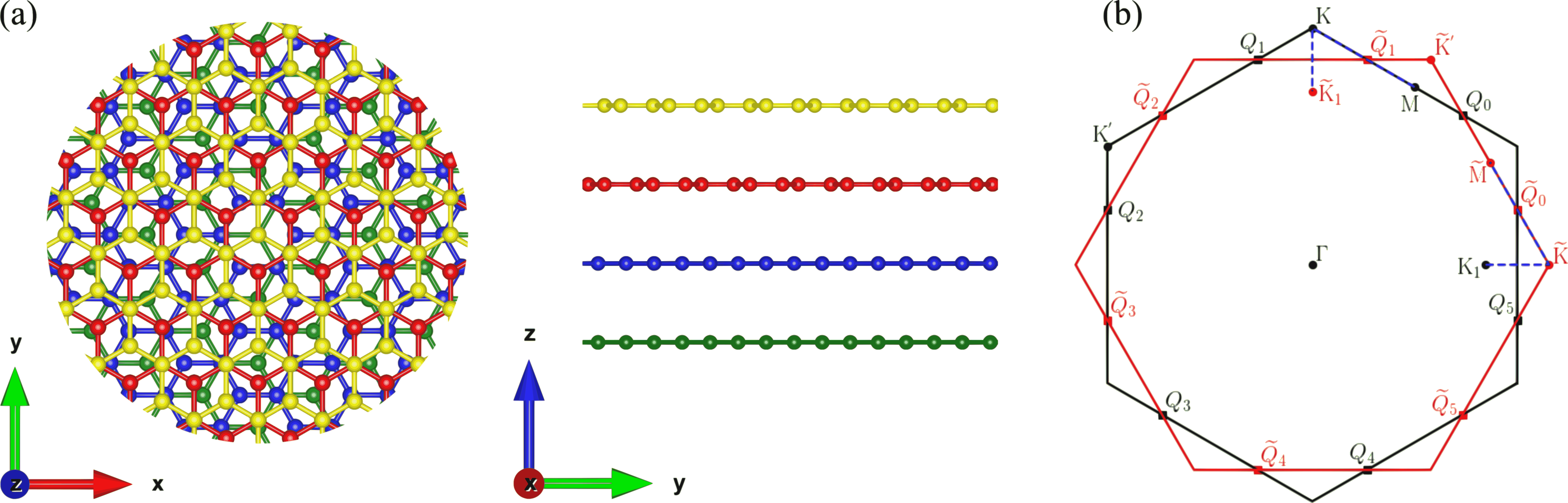}
\caption{(a) The top and side views of 30{\degree} twisted double bilayer graphene. (b) The Brillouin zones of the bottom and top bilayers with some special points labelled.}
\label{fig:struct}
\end{figure*}

\section{methods}

The structure of 30{\degree} TDBG is shown in Fig. \ref{fig:struct}(a). It consists of two AB stacked graphene bilayers with the top bilayer twisted by 30{\degree}. The middle two layers form the 30{\degree} TBG. So there are two AB stacked interfaces and one 30{\degree} twisted interface in 30{\degree} TDBG. In this paper, the 30{\degree} TDBG is approximated by the 15/26 approximant, which is a periodic Moir\'{e} pattern constructed by introducing the slight stress in the top bilayer. The details about the 15/26 approximant are given in Appendix A. 

\begin{figure*}[!htbp]
\centering
\includegraphics[width=16 cm]{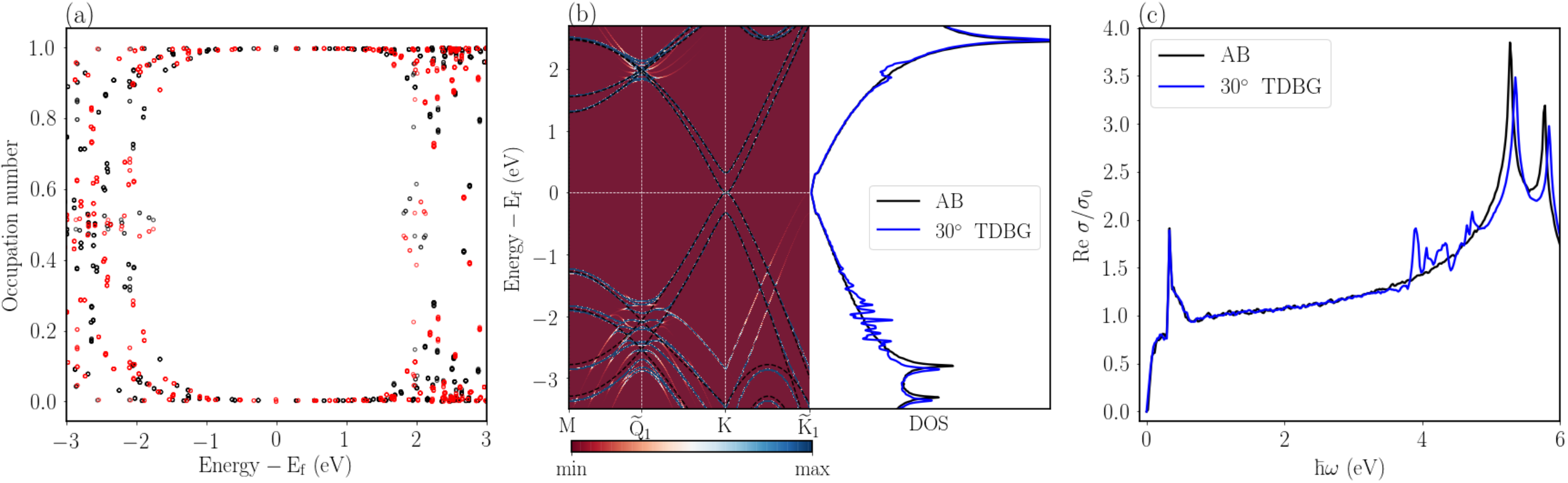}
\caption{(a) The occupation numbers of the eigenstates at $\Gamma$ point. The black and red dots denote the occupation numbers on the bottom and top bilayer, respectively. (b) and (c) The comparisons of the band structures (left part in b), density of states (right part in b) and the real part of the optical conductivity $\sigma$ (c) between 30{\degree} twisted double bilayer graphene and AB stacked bilayer graphene. $\sigma$ is in units of $\sigma_0=\pi e^2/2h$.}
\label{fig:dos_ac_ebs}
\end{figure*}

The 30{\degree} TDBG and its 15/26 approximant are described by the tight-binding model based on $p_z$ orbitals. The hopping energy between site $i$ and $j$ is determined by\cite{LCAO}
\begin{equation}
t_{ij} = n^2 V_{pp\sigma}(\left|\bm{r}_{ij}\right|)+(1-n^2)V_{pp\pi}(\left|\bm{r}_{ij}\right|),
\end{equation}
where, $n$ is the direction cosine of relative position vector $\bm{r}_{ij}$ with respect to the z axis. The Slater-Koster parameters $V_{pp\sigma}$ and $V_{pp\pi}$ have the following forms:
\begin{eqnarray}
V_{pp\pi}(\left|\bm{r}_{ij}\right|)=-\gamma_0 e^{2.218(b-\left|\bm{r}_{ij}\right|)}F_c(\left|\bm{r}_{ij}\right|), \\ 
V_{pp\sigma}(\left|\bm{r}_{ij}\right|)=\gamma_1 e^{2.218(h-\left|\bm{r}_{ij}\right|)}F_c(\left|\bm{r}_{ij}\right|).
\end{eqnarray}
The carbon-carbon distance $b$ and interlayer distance $h$ are chosen to be 3.349 and 1.418 {\AA}, respectively. $\gamma_0$ and $\gamma_1$ are 3.12 and 0.48 eV, respectively. The Fermi velocity and effective band structure calculated using these parameters fit well the experimental results of 30{\degree} TBG\cite{Yu_QC_BG}. $F_c$ is a smooth function
\begin{equation}
F_c(r) = (1+e^{(r-0.265)/5})^{-1}.
\end{equation}
This tight-binding model has also been justified by comparing results with several experiments\cite{TBG_TB_prove1,TBG_TB_prove2,TBG_TB_zhen}. 

The density of states and optical conductivity are calculated by the tight-binding propagation method (TBPM)\cite{TBPM}. This method is based on the numerical solution of time-dependent Schr\"{o}dinger equation without any diagonalization. Both memory and CPU costs scale linearly with the system size. The formula of TBPM are given in Appendix B.

Although the band structure can be derived directly from the periodic 15/26 approximant, its supercell character relative to the primitive unit cell of graphene results in the fold of the bands. So the band structures calculated directly from the 15/26 approximant can not be used to compare with the ARPES measurements. In order to overcome this problem, the band structure of 15/26 approximant is unfolded to the primitive unit cell of graphene. The corresponding formula are given in Appendix C.

\begin{figure}[!htbp]
\centering
\includegraphics[width=8 cm]{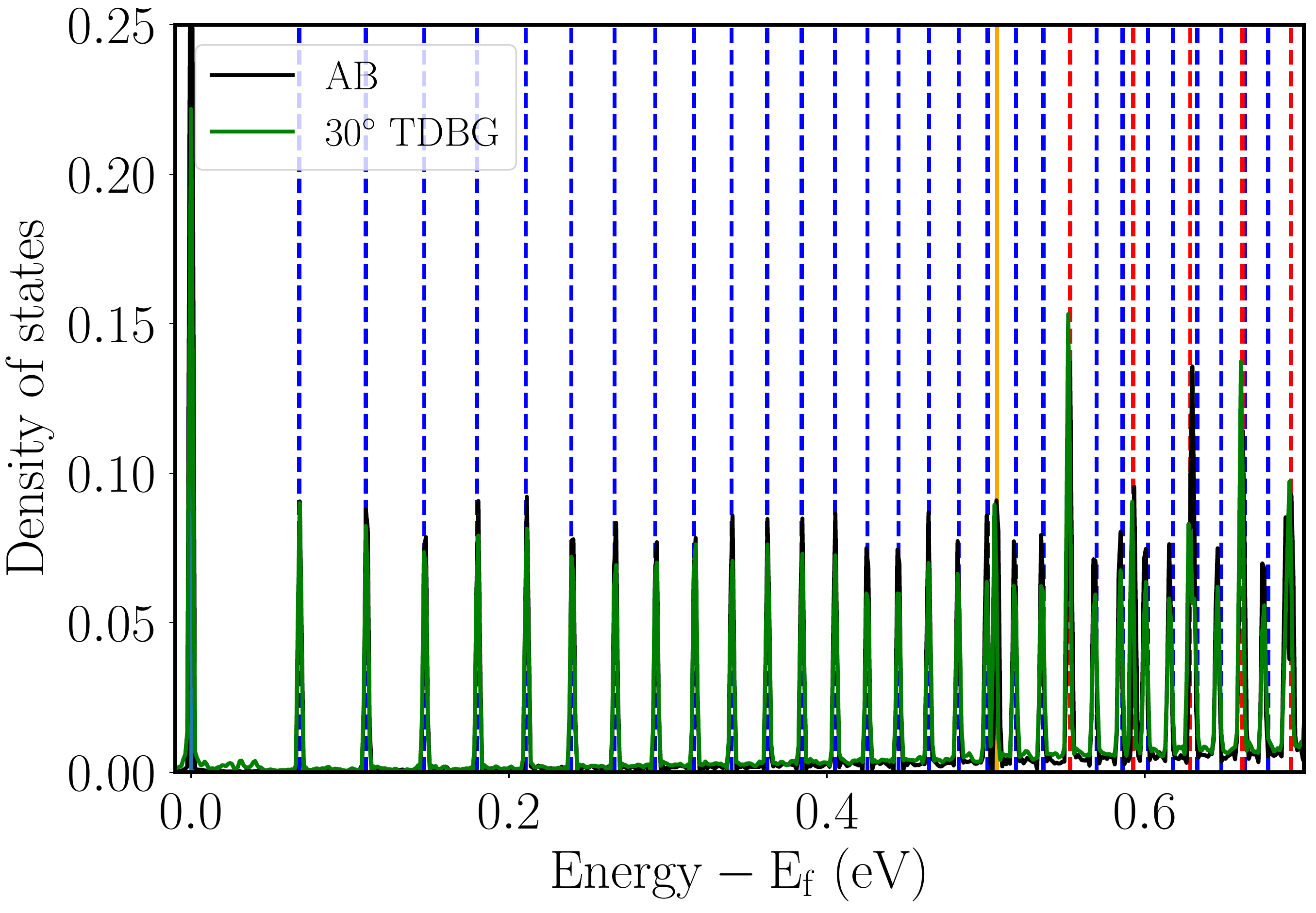}
\caption{The density of states of AB stacked bilayer graphene and 30{\degree} TDBG under the magnetic field of 20 T. The vertical lines show the analytical Landau levels of AB stacked bilayer. The Blue and red dashed lines represent the Landau levels $\varepsilon_{n,L,+}$ and $\varepsilon_{n,H,+}$  with $n \geqslant1$, respectively. The orange solid line is the Landau level $\varepsilon_{0,H,+}$. The levels $\varepsilon_{0,L}$ and $\varepsilon_{-1,L}$ are at zero. Only the electron side are shown here due to the electron-hole symmetry in the simplified tight-binding approximation.}
\label{fig:ll}
\end{figure}

\section{Interlayer decoupling in the low-energy region}
First of all, the distributions of the eigenstates in the low energy region are calculated by diagonalizing the Hamiltonian at $\Gamma$ point and shown in Fig. \ref{fig:dos_ac_ebs}(a). All these states occur inside only the bottom bilayer or top bilayer, which implies the interlayer decoupling between the two bilayer systems in the low energy region. It can be proven further by comparing the band structures and density of states between the 30{\degree} TDBG and its subsystem, namely, the AB stacked bilayer graphene (see Fig. \ref{fig:dos_ac_ebs}(b)). Both of them show the parabolic touch point at K point. The good agreement with each other in the vicinity of the Fermi level means that 30{\degree} TDBG should has the similar electronic properties to the AB stacked bilayer graphene, such as the optical conductivity shown in Fig. \ref{fig:dos_ac_ebs}(c). 

The interlayer decoupling in 30{\degree} TDBG can also be proven by checking the Landau levels (see Fig. \ref{fig:ll}), which is in good agreement with the numerical and analytical\cite{LL_AB} Landau levels of the AB stacked bilayer. In order to compare with the analytical results, a simplified but often used tight-binding approximation is adopted. For the intralayer hoppings, only the nearest-neighbour approximation is adopted. For the interlayer hoppings across the two AB stacked interfaces, only the neighbours stacked vertically are considered. But for the interlayer hoppings across the 30{\degree} twisted interface, all the neighbours with the distance less than 5 {\AA} are taken into account. The analytical Landau levels shown in Fig. \ref{fig:ll} can be classified into three groups.\cite{LL_AB} They are (1) $n\geq1$:
\begin{equation}
\begin{split}
\epsilon_{n,\mu,s} &={{s}\over{\sqrt{2}}}\left[\gamma_1^2 + (2n+1)\Delta_B^2 \right. \\
                   &+ \left. \mu\sqrt{\gamma_1^2 + 2(2n+1)\gamma_1^2\Delta_B^2 + \Delta_B^4} \right]^{1/2},
\end{split} 
\end{equation}
(2)$n=0$: 
\begin{align}
\begin{split} 
\varepsilon_{0,L} &= 0 \\
\varepsilon_{0,H,s} &= s\sqrt{\gamma_1^2+\Delta_B^2},
\end{split}
\end{align}
and (3)$n=-1$:
\begin{equation}
\varepsilon_{-1,L}=0.
\end{equation}
In these equations, $s=\pm1$ (labelled by $\pm$) stand for the electron and hole bands, respectively. In Fig. \ref{fig:ll}, only the electron side is shown due to the electron-hole symmetry in this simplified tight-binding model. $\mu=\pm1$ correspond to the higher and lower subbands in
the limit of zero magnetic field, respectively. In the following, we use the notation $\mu=H,L$ instead of $\pm$ to avoid the confusion with s = $\pm$.    
$\Delta_B$=0.163 eV (defined by $\sqrt{2\hbar v_f^2eB  }$) is the magnetic energy which corresponds to the Fermi velocity $v_f=1\times10^{6}$ m/s in graphene.

Although the interlayer decoupling dominates the electronic properties in the low-energy region, the obvious interlayer coupling still exist in the high-energy region, especially around $\widetilde{Q}_1$ point. Actually, 30{\degree} TDBG shows the similar electronic structure at all $Q$ points ($Q_i$ and $\widetilde{Q}_i$ with i=0, 1, 2, 3, 4 and 5) but deviating from the AB stacked bilayer obviously. It results in the appearance of the new peaks in both density of states and optical conductivity. The origin of the deviation around $Q$ points will be discussed in the next section.        

\begin{figure}[!htbp]
\centering
\includegraphics[width=8.6 cm]{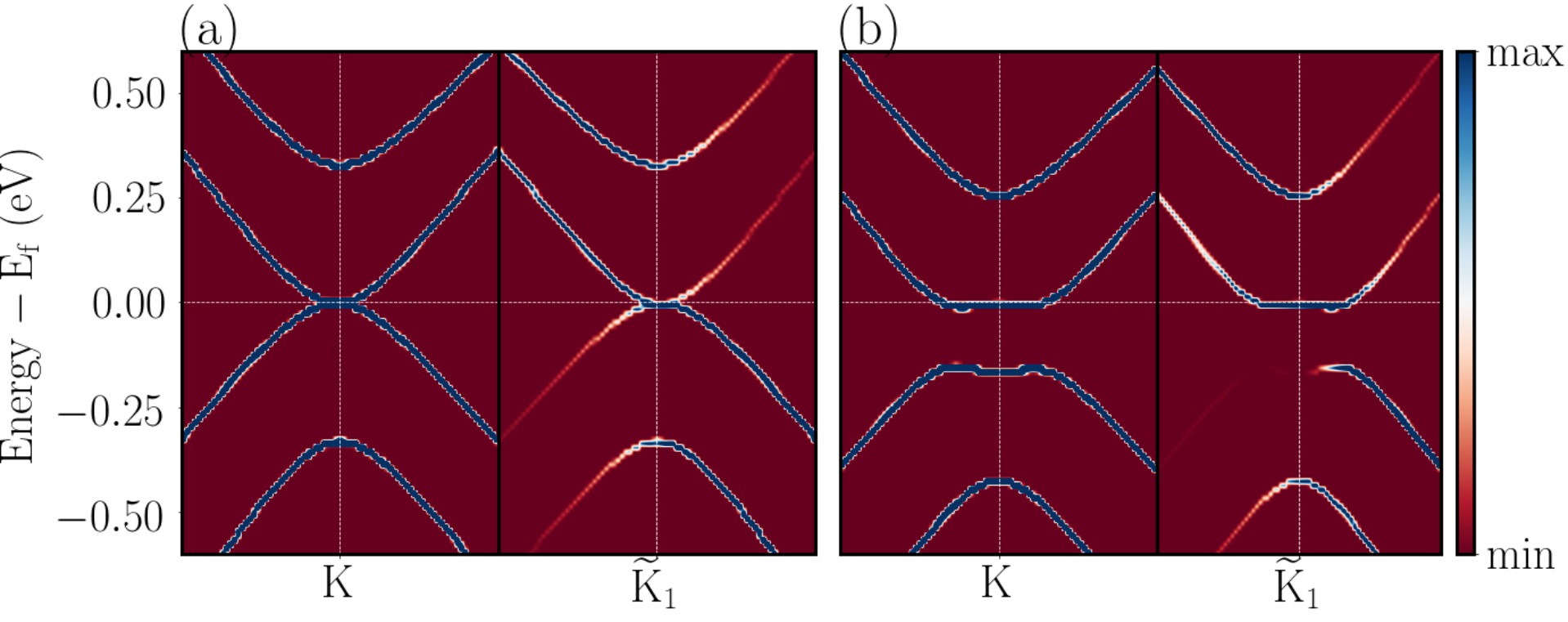}
\caption{The comparison of the effective band structures around $K$ and $\widetilde{K}_1$ along $x$ direction under the electric field of 0 (a) and 0.05 (b) eV/{\AA}. The spectral function values around $\widetilde{K}_1$ are timed by 300 for clear comparison.}
\label{fig:ebs_onek}
\end{figure}

\begin{figure*}[!htbp]
\centering
\includegraphics[width=14 cm]{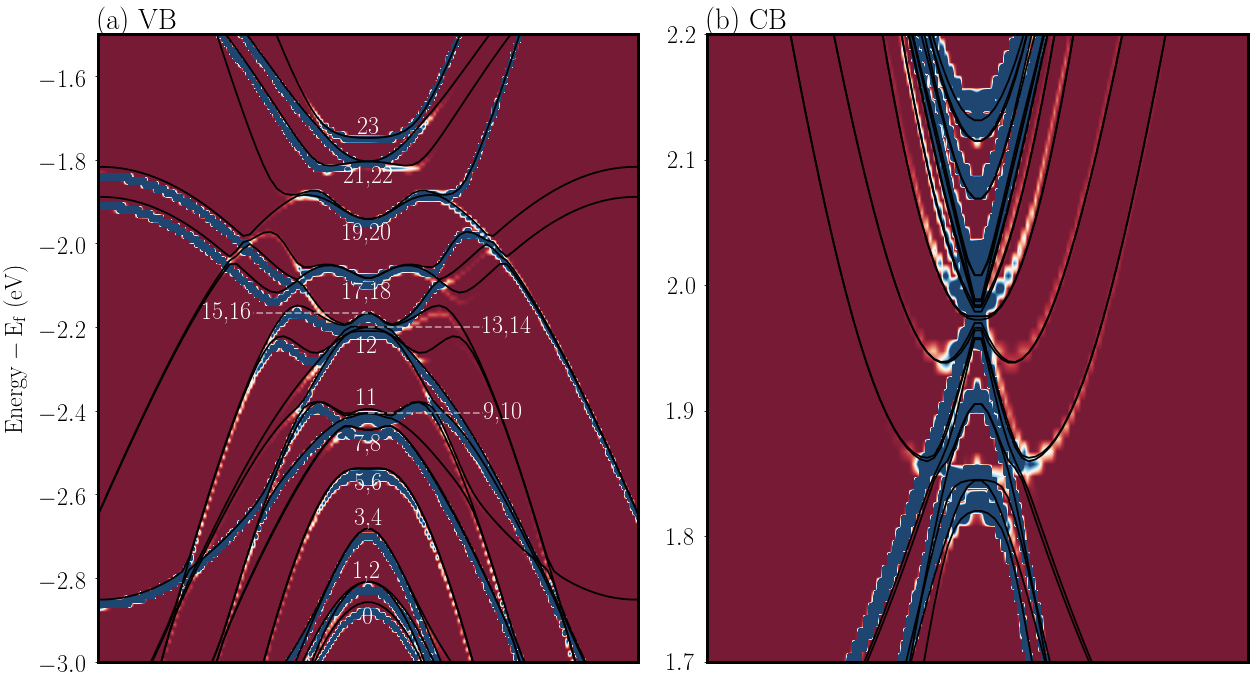}
\caption{The comparisons of the effective band structure around $\widetilde{Q}_1$ (plotted in color) and the quasi-band structure around $k_0=0$ (plotted in black solid lines) in the valence band (a) and the conduction band (b). In valence band, except for the states 0, 11, 12 and 23, any two states marked by i,i+1, are energy degenerate.}
\label{fig:ebs_qbs}
\end{figure*}
Similar to 30{\degree} TBG\cite{pnas_QC,science_QC}, another important property is the emergence of the energy valley at $\widetilde{K}_1$, which is mirror-symmetric with respect to $K$ point (see Fig. \ref{fig:ebs_onek}(a)) because the energy valley at $K^{'}$ is scattered to $\widetilde{K}_1$ with a strong scattering strength $\left| \braket{\widetilde{K}_1,\widetilde{X}|U|K^{'},X} \right|$. $U$ is the interaction between the bottom and top bilayers. $\left|K^{'},X \right>$ and $\left| \widetilde{K}_1, \widetilde{X} \right>$ are the Bloch functions of the bottom and top bilayer systems, respectively. The Bloch function of AB stacked bilayer graphene at a general point $\bm k$ and sublattice $X$ is defined by
\begin{equation}
\left| \bm{k},X \right> = {{1}\over{\sqrt{n}}} \sum_{\bm{R}}  e^{i\bm{k}\cdot(\bm{R} + \bm{\tau_X})} \left| \varphi(\bm{r}-\bm{R}-\bm{\tau}_X) \right>.
\end{equation}
Here, $n$ is the normalization factor, $\bm{\tau}_X$ is the position of sublattice $X$, and $\varphi(\bm{r}-\bm{R}-\bm{\tau}_X)$ is the $p_z$ orbital locating at sublattice $X$ in unit cell $\bm{R}$. The interlayer decoupling and energy valley scattering are robust even an electric field is applied perpendicular to the graphene plane (see Fig. \ref{fig:ebs_onek}(b)).

\section{Interlayer coupling at $Q$ points}

\begin{figure*}[!htbp]
\centering
\includegraphics[width=18 cm]{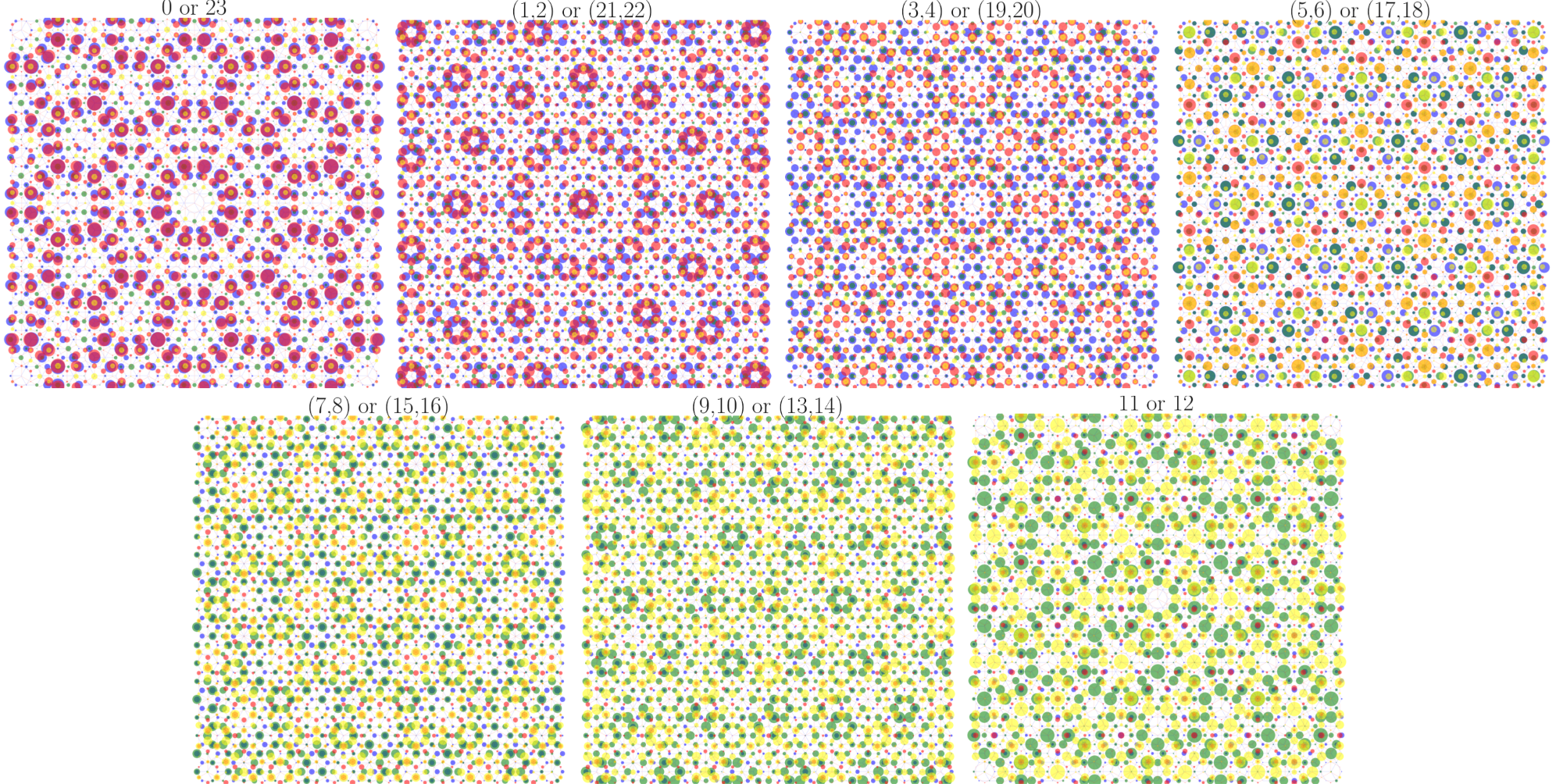}
\caption{The charge distributions of the electronic states at $\widetilde{Q}_1$ point in the valence band. The labels of these states are marked in Fig.\ref{fig:ebs_qbs}(a). The occupations on the four layers from bottom to top are plotted in green, blue, red and yellow dots.  i or j means states i and j have the same charge distribution pattern. (i,j) corresponds to the charge distribution of the 2-fold degenerate states i and j being occupied at the same time.}
\label{fig:wfs}
\end{figure*}

The effective band structure of 30{\degree} TDBG deviates obviously from the band structure of AB stacked bilayer graphene around $Q$ points (see Fig. \ref{fig:dos_ac_ebs}(b)), which implies the strong interlayer coupling across the 30{\degree} twisted interface. The k-space tight-binding method\cite{arXiv_QC} is adopted to understand this phenomenon. In this method, a $\bm{k}_0$-related subspace is spanned by Bloch basis functions of both the top and bottom bilayers, namely $\lbrace\left|\bm{k}_0 + \bm{G}, \widetilde{X}\right>\rbrace$, and $\lbrace\left|\bm{k}_0 + \bm{\widetilde{G}}, X\right>\rbrace$, where $\bm{G}$ ($X$) and $\bm{\widetilde{G}}$ ($\widetilde{X}$) are the reciprocal lattice vectors (sublattice) of the bottom and top bilayers, respectively. Actually, only the $\bm{G}$'s and $\bm{\widetilde{G}}$'s with small lengths contribute the Hamiltonian much. In this paper, the 12-wave approximation\cite{arXiv_QC}, namely only considering the $\bm G$'s and $\bm{\widetilde{G}}$'s with the length less than ${{4\pi}\over{\sqrt{3}a}}$, is adopted to construct the Hamiltonian, which has been proven to be accurate enough to simulate the 30{\degree} TBG.\cite{arXiv_QC} The matrix element of the Hamiltonian across the different bilayer systems is
\begin{equation}
\begin{split}
\braket{\bm{k}_0+\bm{\widetilde{G}}, X|U  |\bm{k}_0+\bm{G}, \widetilde{X}} = \\
 T(\bm{k}_0+\bm{G}+\bm{\widetilde{G}})e^{-i\bm{\widetilde{G}}\cdot\bm{\tau}_{\widetilde{X}}}e^{i\bm{G}\cdot{\bm{\tau}_{X}}},
\end{split}
\end{equation}
where $T(\bm{k}_0+\bm{G}+\bm{\widetilde{G}})$ is the Fourier component of the interlayer hopping function at vector $\bm{k}_0+\bm{G}+\bm{\widetilde{G}}$.  

Around $\bm k_0$=0 and under the 12-wave approximation, after folding $\bm{k}_0+\bm{\widetilde{G}}$ of the bottom bilayer and $\bm{k}_0+\bm{G}$ of the top bilayer to their corresponding first Brillouin zones, the basis set is just the collection of the Bloch functions of the bottom layer $\lbrace \left| Q_i, X \right> \rbrace$ and the Bloch functions of the top bilayer $\lbrace \left| \widetilde{Q}_i, \widetilde{X} \right> \rbrace$ with i=0, 1, 2 3, 4 and 5 (see Fig. \ref{fig:struct}(b) for their positions). Because there are four sublattices in each bilayer system, the Hamiltonian is a $48\times48$ matrix. After the diagonalization of the Hamiltonian, the dispersion relationship between the energy and $\bm k_0$ can be obtained, which is named as quasi-band structure to distinguish the term effective band structure derived by unfolding the band structure of the 15/26 approximant. Due to the energy degeneration of the Bloch states at the 12 $Q$ points, the strong interactions among them results in the strong deviation of the effective band structure of the 30{\degree} TDBG from AB stacked bilayer graphene. In Fig. \ref{fig:ebs_qbs}, we plot the quasi-band structure around $\bm k_0$=0 and compare it with the effective band structure around $\widetilde{Q}_1$ obtained by band-unfolding method. The good agreement with each other proves the validations of the 12-wave approximation and the 15/26 approximant again. Due to the weaker interaction in the conduction band, we only focus on the electronic structure in the valence band. 

At $\bm k_0$=0, the 24 electronic states in the valence band are labelled by 0, ..., 23 (see Fig. \ref{fig:ebs_qbs}(a)). Expect for states 0, 11, 12 and 23, the others are all 2-fold degenerate states. The charge distributions of these states are shown in Fig. \ref{fig:wfs}, where (i,j) corresponds to the charge distribution of the 2-fold degenerate states i and j being occupied at the same time. (i,j) or (m,n) means the two charge distributions have the similar pattern and are plotted in the same sub-figure. Comparing with 30{\degree} TBG with the symmetry of point group $D_{6d}$, all the charge distribution patterns in 30{\degree} TDBG lose the 12-fold symmetry due to the symmetry decrease to point group $D_3$. But the corresponding 12-fold-symmetry-like counterparts still exist in 30{\degree} TDBG. For example, the occupation number of the charge distribution 0 on the middle two layers is more than 85\%. If the occupation on the bottom and top layers is ignored, this charge distribution pattern still possess the 12-fold symmetry, which correspond to the charge distribution $m=0$ or $6$ in 30{\degree} TBG\cite{arXiv_QC}. Besides, the charge distributions (1,2) and (3,4) have the occupation number on the middle two layers more than 70\%, correspond to the cases $m=\pm1, \pm5$ and $m=\pm2, \pm4$ in 30{\degree} TBG\cite{arXiv_QC}, respectively. The charge distribution (5,6) is similar to the case $m=\pm3$ in 30{\degree} TBG\cite{arXiv_QC}, but they show the similar occupation numbers on the four layers. Importantly, all the charge distributions (7,8), (9,10) and 11 show very different charge distribution patterns from the 30{\degree} TBG, and these charge distributions have very little occupation number on the middle two layers (less than 20\%). These results indicate that the 12-wave interaction results in the strong interlayer coupling between the two bilayer systems at $\bm k_0=0$, namely at all $Q$ points.

\section{Discussion}
\begin{figure}[!htbp]
\centering
\includegraphics[width=7 cm]{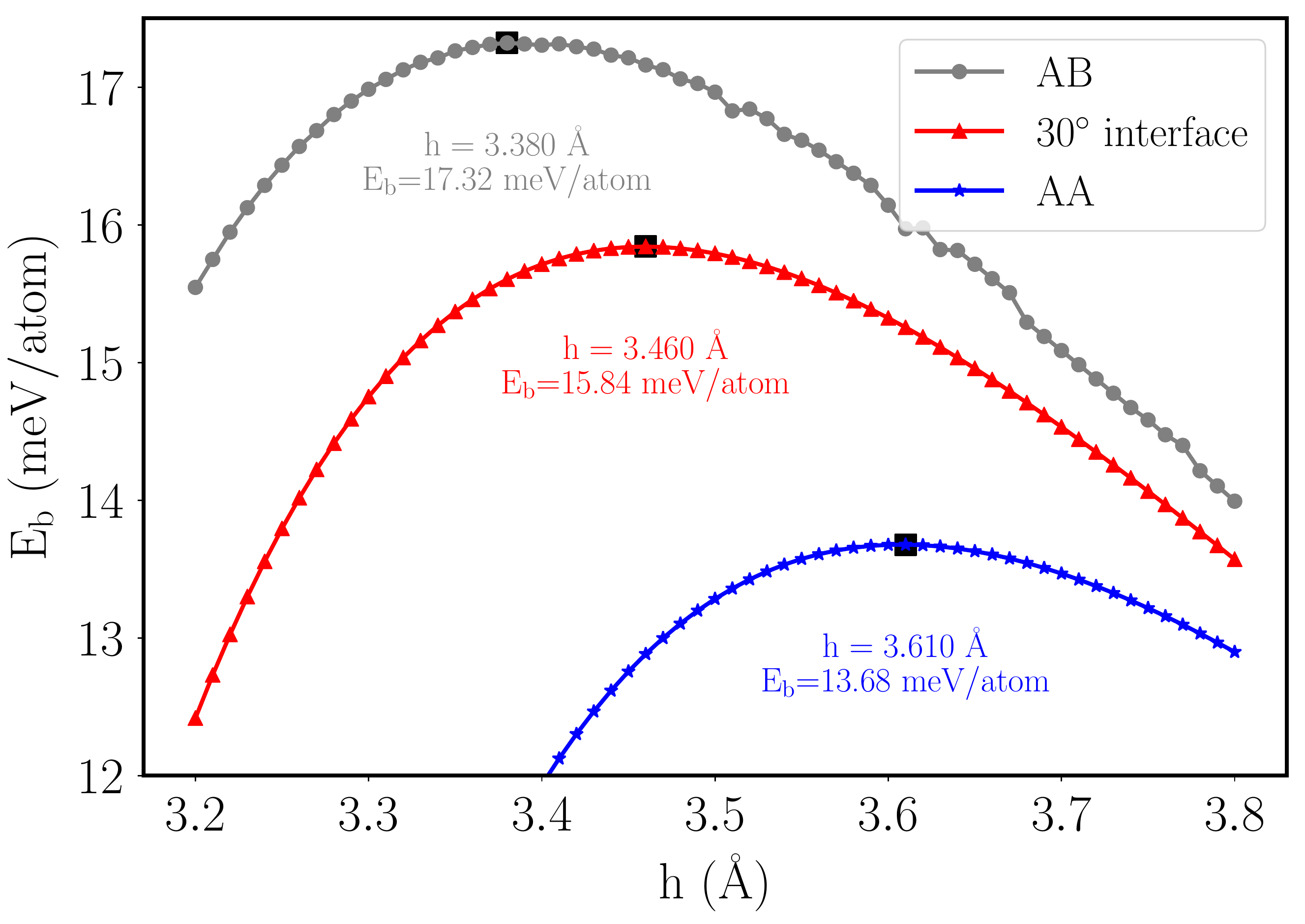}
\caption{The curves of the binding energy with respect to the interlayer spacing in AA, AB stacked and 30{\degree} twisted interfaces. The binding energy and interlayer spacing at the most stable configure is given for each interface.}
\label{fig:Eb}
\end{figure}

During the calculations above, the interlayer distances in both the AB stacked and 30{\degree} twisted interfaces are fixed to be 3.349 {\AA}, which may not be their most stable configurations. In order to consider the structure relaxation effect, the interlayer distances in AA, AB stacked and 30{\degree} twisted interfaces are determined by classical molucular simulations. The intra- and inter-layer interactions are described by classical REBO\cite{reb0} and Kolmogorov-Crespi\cite{KC_potential} potentials, which are implemented in the LAMMPS software\cite{lammps}. For a fixed interlayer distance, the atoms are allowed to relax only in $xy$ plane. The binding energies of the three interfaces at different interlayer distances are given in Fig. \ref{fig:Eb}. At the most stable configurations, the interlayer distances are determined to be 3.61, 3.38 and 3.46 {\AA} for AA, AB stacked and 30{\degree} twisted interfaces, respectively, which correspond to the binding energies 13.7, 17.3 and 15.8 meV/atom. Different from the small twist angle, for which the fluctuations of the atoms along the $z$ axis introduce the periodic pseudo magnetic field,\citep{TBG_TB_zhen} the flatness of the two layers still remains in the 30{\degree} twisted interface. After the recalculations by using the relaxed interlayer distances, all the physical conclusions concluded above still be kept.

From the viewpoint of the binding energy, the 30{\degree} twisted interface is much more stable than the AA stacked interface and even comparable to the AB stacked one. It means that the 30{\degree} twisted interface can be realized easily in experiment. So we also study the effect of the 30{\degree} twisted interface between AB stacked bilayer and graphene monolayer (labelled by 30{\degree} AB/G), which can be obtained by removing the top layer of the 30{\degree} TDBG. Our calculations shown in the Supplemental Material\cite{sm} indicate that the interlayer decoupling in the low-energy region and the strong interlayer coupling at $Q$ points still exist in this system, which implies that it is a general phenomenon in graphene multilayer systems.
        
\section{Conclusion}
By means of the tight-binding approximation, we systematically study the electronic properties of 30{\degree} TDBG, which are composed of two AB stacked bilayer graphene with top bilayer twisted by 30{\degree}. In the low-energy region, the interlayer decoupling across the 30{\degree} twisted interface is proven from various electronic properties, such as density of states, effective band structure, optical conductivity and Landau levels. However, the 30{\degree} TDBG shows very different effective band structure at $Q$ points from the AB stacked bilayer due to the 12-wave interaction, which results in the appearance of new van Hove singularities in the density of states and new peaks in the optical conductivity. Importantly, the 12-fold-symmetry-like electronic states, which occur in 30{\degree} TBG with exact 12-fold symmetry, can be found in 30{\degree} TDBG due to the 12-wave interaction, although its symmetry decreases to point group $D_3$. Moreover, different from the 30{\degree} TBG, some special electronic states appear, which have large occupation number on the top and bottom layers but little occupation nunber on the middle two layers. These results imply that the strong interlayer coupling still exist across the 30{\degree} twist in 30{\degree} TDBG, especially at $Q$ points, although it shows interlayer decoupling in the vicinity of the Fermi level.

\section*{ACKNOWLEDGEMENTS}
This work is supported by the National Science Foundation of China (Grant No. 11774269) and China Postdoctoral Science Foundation (Grant No. 2018M632902). MIK acknowledges a support by the JTC-FLAGERA Project GRANSPORT. Numerical calculations presented in this paper have been partially performed on the supercomputing system in the Supercomputing Center of Wuhan University. Support by the Netherlands National Computing Facilities foundation (NCF), with funding from the Netherlands Organisation for Scientific Research (NWO), is gratefully acknowledged.

\begin{appendices}
\section{15/26 approximant}
We follow the procedure proposed for 30{\degree} TBG\cite{Yu_QC_BG} to construct the approximant of 30{\degree} TDBG. That is, the bottom bilayer keeps the lattice constant of a pristine graphene $a=2.456$ {\AA}, but the top bilayer is slightly compressed with the lattice constant changing to be $\widetilde{a}=2.454$ {\AA}, which makes the two bilayers become commensurate. The resulting periodic pattern is used as the approximant, which is named as 15/26 approximant due to the commensurate period $15\times \sqrt{3}a=26\times \widetilde{a}$ along the $x$ direction, where $\sqrt{3}a$ and $\widetilde{a}$ are the basic periods of the bottom and top bilayers, respectively. Accordingly, the elementary unit cell of the 15/26 approximant contains 1350$\times$2 and 1352$\times$2 sites in the bottom and top bilayers, respectively. Our results given in the Supplemental Material\cite{sm} indicate that the 15/26 approximant can reproduce the density of states and the optical conductivity of 30{\degree} TDBG accurately. 

\section{Tight-binding propagation method}
In TBPM\cite{TBPM}, a random superposition of the $p_z$ orbitals at all sites is used as the initial state $\left|\phi_0 \right>$ with $\braket{\phi_0|\phi_0}=1$. DOS is calculated as Fourier transform of the time-dependent correlation function
\begin{equation}
d(\epsilon)={{1}\over{2\pi}}\int_{-\infty}^\infty e^{i\epsilon\tau}\braket{\phi_0|e^{-iH\tau/\hbar}|\phi_0}d\tau. 
\end{equation}

The optical conductivity is calculated by using the Kubo formula in TBPM\cite{TBPM}. The real part of the optical conductivity matrix $\sigma_{\alpha,\beta}$ at temperature $T$ reads
\begin{equation}
\begin{aligned}
Re\sigma_{\alpha,\beta}(\omega) = \lim_{\epsilon\rightarrow 0^+} {{e^{-\hbar\omega/k_BT}-1}\over{\hbar \omega A}}\int^\infty_0 e^{-\epsilon\tau}sin\omega\tau \\ \times 2Im\braket{\phi_2(\tau)|j_\alpha|\phi_1(\tau)}_\beta d\tau.
\end{aligned}
\end{equation}
Here, $A$ is the area of the unit cell per layer, and wave functions
\begin{equation}
\begin{aligned}
\left| \phi_1(\tau)\right>_\beta = e^{-iH\tau/\hbar}[1-f(H)]j_\beta\left| \phi_0\right>, \\
\left| \phi_2(\tau) \right> = e^{-iH\tau/\hbar} f(H)\left| \phi_0 \right>,
\end{aligned}
\end{equation}
where $f(H)=1/(e^{{{(H-\mu)}/{k_BT}}}+1)$ is the Fermi-Dirac distribution operator, where $\mu$ is the electronic chemical potential. 

\section{effective band structure}
First of all, the spectral function at wavevector $\bm{k}$ and energy $\epsilon$ can be calculated by\cite{BandUnfold_SF}
\begin{equation}
A(\bm{k},\epsilon) = \sum_{I\bm{k}_{SC}}P_{I\bm{k}_{SC}}(\bm{k})\delta(\epsilon-\epsilon_{I\bm{k}_{SC}}),
\end{equation}      
where $\epsilon_{I\bm{k}_{SC}}$ is the energy for $I^{th}$ band at wavevector $\bm{k}_{SC}$ for the approximant. Actually, only one $\bm{k}_{SC}$, namely $\bm{k}_{SC} = \bm{k} + \bm{G}$ being $\bm{G}$ the reciprocal lattice vector of the approximant, contributes to the spectral function. 
The spectral weight is defined as
\begin{equation}
P_{I\bm{k}_{SC}}(\bm{k})=\sum_{s} \sum_i \left|  \braket{\psi^{PC_s}_{i\bm{k}}| \Psi^{SC}_{I\bm{k}_{SC}}}  \right|^2=\sum_{s} P_{I\bm{k}_{SC}}^s(\bm{k}),
\end{equation}
where $\left| \psi^{PC_s}_{i\bm{k}}\right>$  and $\left|\Psi^{SC}_{I\bm{k}_{SC}} \right>$ are the eigenstates of layer $s$ and the approximant, respectively. Under the tight-binding method, the spectral weight contributed from layer $s$ can be described by
\begin{equation}\label{spec_weig}
P_{I\bm{k}_{SC}}^s(\bm{k})= {{1}\over{n_s}}\sum_\alpha \sum_{\bm{l_s}\bm{l_s}^{'}}e^{i\bm{k}\cdot(\bm{l_s}-\bm{l_s}^{'})}U^{\bm{l_s}\alpha^*}_{I\bm{k}_{SC}} U^{\bm{l_s}^{'}\alpha}_{I\bm{k}_{SC}}.
\end{equation}
Here, $n_s$ is the number of primitive unit cell of layer $s$ in one elementary unit cell of the approximant. $U^{\bm{l_s}\alpha}_{I\bm{k}_{SC}}$ is the projection of $\left|\Psi^{SC}_{I\bm{k}_{SC}} \right>$ (the eigenstate of the approximant) on $\left| \bm{k}_{SC} \bm{l_s}\alpha\right>$ (the Bloch basis function of approximant). Equation (\ref{spec_weig}) indicates that only the eigenstates of approximant are necessary to obtain the spectral function. 

Then, the effective band structure can be obtained by\cite{BandUnfold_EBS}
\begin{equation}
\delta N(\bm{k},\epsilon) = \int_{\epsilon-\delta \epsilon/2} ^ {\epsilon+\delta \epsilon/2} A(\bm{k}, \epsilon^{'}) d \epsilon^{'},
\end{equation}
where $\delta \epsilon$ is the bin width in energy sampling.
\end{appendices}

\bibliography{30_tMG_QC.bib}

\begin{thebibliography}{55}%
\makeatletter
\providecommand \@ifxundefined [1]{%
 \@ifx{#1\undefined}
}%
\providecommand \@ifnum [1]{%
 \ifnum #1\expandafter \@firstoftwo
 \else \expandafter \@secondoftwo
 \fi
}%
\providecommand \@ifx [1]{%
 \ifx #1\expandafter \@firstoftwo
 \else \expandafter \@secondoftwo
 \fi
}%
\providecommand \natexlab [1]{#1}%
\providecommand \enquote  [1]{``#1''}%
\providecommand \bibnamefont  [1]{#1}%
\providecommand \bibfnamefont [1]{#1}%
\providecommand \citenamefont [1]{#1}%
\providecommand \href@noop [0]{\@secondoftwo}%
\providecommand \href [0]{\begingroup \@sanitize@url \@href}%
\providecommand \@href[1]{\@@startlink{#1}\@@href}%
\providecommand \@@href[1]{\endgroup#1\@@endlink}%
\providecommand \@sanitize@url [0]{\catcode `\\12\catcode `\$12\catcode
  `\&12\catcode `\#12\catcode `\^12\catcode `\_12\catcode `\%12\relax}%
\providecommand \@@startlink[1]{}%
\providecommand \@@endlink[0]{}%
\providecommand \url  [0]{\begingroup\@sanitize@url \@url }%
\providecommand \@url [1]{\endgroup\@href {#1}{\urlprefix }}%
\providecommand \urlprefix  [0]{URL }%
\providecommand \Eprint [0]{\href }%
\providecommand \doibase [0]{http://dx.doi.org/}%
\providecommand \selectlanguage [0]{\@gobble}%
\providecommand \bibinfo  [0]{\@secondoftwo}%
\providecommand \bibfield  [0]{\@secondoftwo}%
\providecommand \translation [1]{[#1]}%
\providecommand \BibitemOpen [0]{}%
\providecommand \bibitemStop [0]{}%
\providecommand \bibitemNoStop [0]{.\EOS\space}%
\providecommand \EOS [0]{\spacefactor3000\relax}%
\providecommand \BibitemShut  [1]{\csname bibitem#1\endcsname}%
\let\auto@bib@innerbib\@empty
%</preamble>
\bibitem [{\citenamefont {Shallcross}\ \emph {et~al.}(2010)\citenamefont
  {Shallcross}, \citenamefont {Sharma}, \citenamefont {Kandelaki},\ and\
  \citenamefont {Pankratov}}]{commensuration_condition}%
  \BibitemOpen
  \bibfield  {author} {\bibinfo {author} {\bibfnamefont {S.}~\bibnamefont
  {Shallcross}}, \bibinfo {author} {\bibfnamefont {S.}~\bibnamefont {Sharma}},
  \bibinfo {author} {\bibfnamefont {E.}~\bibnamefont {Kandelaki}}, \ and\
  \bibinfo {author} {\bibfnamefont {O.~A.}\ \bibnamefont {Pankratov}},\ }\href
  {\doibase 10.1103/PhysRevB.81.165105} {\bibfield  {journal} {\bibinfo
  {journal} {Phys. Rev. B}\ }\textbf {\bibinfo {volume} {81}},\ \bibinfo
  {pages} {165105} (\bibinfo {year} {2010})}\BibitemShut {NoStop}%
\bibitem [{\citenamefont {Berger}\ \emph {et~al.}(2006)\citenamefont {Berger},
  \citenamefont {Song}, \citenamefont {Li}, \citenamefont {Wu}, \citenamefont
  {Brown}, \citenamefont {Naud}, \citenamefont {Mayou}, \citenamefont {Li},
  \citenamefont {Hass}, \citenamefont {Marchenkov}, \citenamefont {Conrad},
  \citenamefont {First},\ and\ \citenamefont {de~Heer}}]{tBG_15plus0}%
  \BibitemOpen
  \bibfield  {author} {\bibinfo {author} {\bibfnamefont {C.}~\bibnamefont
  {Berger}}, \bibinfo {author} {\bibfnamefont {Z.}~\bibnamefont {Song}},
  \bibinfo {author} {\bibfnamefont {X.}~\bibnamefont {Li}}, \bibinfo {author}
  {\bibfnamefont {X.}~\bibnamefont {Wu}}, \bibinfo {author} {\bibfnamefont
  {N.}~\bibnamefont {Brown}}, \bibinfo {author} {\bibfnamefont
  {C.}~\bibnamefont {Naud}}, \bibinfo {author} {\bibfnamefont {D.}~\bibnamefont
  {Mayou}}, \bibinfo {author} {\bibfnamefont {T.}~\bibnamefont {Li}}, \bibinfo
  {author} {\bibfnamefont {J.}~\bibnamefont {Hass}}, \bibinfo {author}
  {\bibfnamefont {A.~N.}\ \bibnamefont {Marchenkov}}, \bibinfo {author}
  {\bibfnamefont {E.~H.}\ \bibnamefont {Conrad}}, \bibinfo {author}
  {\bibfnamefont {P.~N.}\ \bibnamefont {First}}, \ and\ \bibinfo {author}
  {\bibfnamefont {W.~A.}\ \bibnamefont {de~Heer}},\ }\href {\doibase
  10.1126/science.1125925} {\bibfield  {journal} {\bibinfo  {journal}
  {Science}\ }\textbf {\bibinfo {volume} {312}},\ \bibinfo {pages} {1191}
  (\bibinfo {year} {2006})}\BibitemShut {NoStop}%
\bibitem [{\citenamefont {Hass}\ \emph {et~al.}(2008)\citenamefont {Hass},
  \citenamefont {Varchon}, \citenamefont {Mill\'an-Otoya}, \citenamefont
  {Sprinkle}, \citenamefont {Sharma}, \citenamefont {de~Heer}, \citenamefont
  {Berger}, \citenamefont {First}, \citenamefont {Magaud},\ and\ \citenamefont
  {Conrad}}]{tBG_15plus1}%
  \BibitemOpen
  \bibfield  {author} {\bibinfo {author} {\bibfnamefont {J.}~\bibnamefont
  {Hass}}, \bibinfo {author} {\bibfnamefont {F.}~\bibnamefont {Varchon}},
  \bibinfo {author} {\bibfnamefont {J.~E.}\ \bibnamefont {Mill\'an-Otoya}},
  \bibinfo {author} {\bibfnamefont {M.}~\bibnamefont {Sprinkle}}, \bibinfo
  {author} {\bibfnamefont {N.}~\bibnamefont {Sharma}}, \bibinfo {author}
  {\bibfnamefont {W.~A.}\ \bibnamefont {de~Heer}}, \bibinfo {author}
  {\bibfnamefont {C.}~\bibnamefont {Berger}}, \bibinfo {author} {\bibfnamefont
  {P.~N.}\ \bibnamefont {First}}, \bibinfo {author} {\bibfnamefont
  {L.}~\bibnamefont {Magaud}}, \ and\ \bibinfo {author} {\bibfnamefont {E.~H.}\
  \bibnamefont {Conrad}},\ }\href {\doibase 10.1103/PhysRevLett.100.125504}
  {\bibfield  {journal} {\bibinfo  {journal} {Phys. Rev. Lett.}\ }\textbf
  {\bibinfo {volume} {100}},\ \bibinfo {pages} {125504} (\bibinfo {year}
  {2008})}\BibitemShut {NoStop}%
\bibitem [{\citenamefont {Miller}\ \emph {et~al.}(2009)\citenamefont {Miller},
  \citenamefont {Kubista}, \citenamefont {Rutter}, \citenamefont {Ruan},
  \citenamefont {de~Heer}, \citenamefont {First},\ and\ \citenamefont
  {Stroscio}}]{tBG_15plus2}%
  \BibitemOpen
  \bibfield  {author} {\bibinfo {author} {\bibfnamefont {D.~L.}\ \bibnamefont
  {Miller}}, \bibinfo {author} {\bibfnamefont {K.~D.}\ \bibnamefont {Kubista}},
  \bibinfo {author} {\bibfnamefont {G.~M.}\ \bibnamefont {Rutter}}, \bibinfo
  {author} {\bibfnamefont {M.}~\bibnamefont {Ruan}}, \bibinfo {author}
  {\bibfnamefont {W.~A.}\ \bibnamefont {de~Heer}}, \bibinfo {author}
  {\bibfnamefont {P.~N.}\ \bibnamefont {First}}, \ and\ \bibinfo {author}
  {\bibfnamefont {J.~A.}\ \bibnamefont {Stroscio}},\ }\href {\doibase
  10.1126/science.1171810} {\bibfield  {journal} {\bibinfo  {journal}
  {Science}\ }\textbf {\bibinfo {volume} {324}},\ \bibinfo {pages} {924}
  (\bibinfo {year} {2009})}\BibitemShut {NoStop}%
\bibitem [{\citenamefont {Sadowski}\ \emph {et~al.}(2006)\citenamefont
  {Sadowski}, \citenamefont {Martinez}, \citenamefont {Potemski}, \citenamefont
  {Berger},\ and\ \citenamefont {de~Heer}}]{tBG_15plus3}%
  \BibitemOpen
  \bibfield  {author} {\bibinfo {author} {\bibfnamefont {M.~L.}\ \bibnamefont
  {Sadowski}}, \bibinfo {author} {\bibfnamefont {G.}~\bibnamefont {Martinez}},
  \bibinfo {author} {\bibfnamefont {M.}~\bibnamefont {Potemski}}, \bibinfo
  {author} {\bibfnamefont {C.}~\bibnamefont {Berger}}, \ and\ \bibinfo {author}
  {\bibfnamefont {W.~A.}\ \bibnamefont {de~Heer}},\ }\href {\doibase
  10.1103/PhysRevLett.97.266405} {\bibfield  {journal} {\bibinfo  {journal}
  {Phys. Rev. Lett.}\ }\textbf {\bibinfo {volume} {97}},\ \bibinfo {pages}
  {266405} (\bibinfo {year} {2006})}\BibitemShut {NoStop}%
\bibitem [{\citenamefont {Sprinkle}\ \emph {et~al.}(2009)\citenamefont
  {Sprinkle}, \citenamefont {Siegel}, \citenamefont {Hu}, \citenamefont
  {Hicks}, \citenamefont {Tejeda}, \citenamefont {Taleb-Ibrahimi},
  \citenamefont {Le~F\`evre}, \citenamefont {Bertran}, \citenamefont {Vizzini},
  \citenamefont {Enriquez}, \citenamefont {Chiang}, \citenamefont
  {Soukiassian}, \citenamefont {Berger}, \citenamefont {de~Heer}, \citenamefont
  {Lanzara},\ and\ \citenamefont {Conrad}}]{tBG_15plus4}%
  \BibitemOpen
  \bibfield  {author} {\bibinfo {author} {\bibfnamefont {M.}~\bibnamefont
  {Sprinkle}}, \bibinfo {author} {\bibfnamefont {D.}~\bibnamefont {Siegel}},
  \bibinfo {author} {\bibfnamefont {Y.}~\bibnamefont {Hu}}, \bibinfo {author}
  {\bibfnamefont {J.}~\bibnamefont {Hicks}}, \bibinfo {author} {\bibfnamefont
  {A.}~\bibnamefont {Tejeda}}, \bibinfo {author} {\bibfnamefont
  {A.}~\bibnamefont {Taleb-Ibrahimi}}, \bibinfo {author} {\bibfnamefont
  {P.}~\bibnamefont {Le~F\`evre}}, \bibinfo {author} {\bibfnamefont
  {F.}~\bibnamefont {Bertran}}, \bibinfo {author} {\bibfnamefont
  {S.}~\bibnamefont {Vizzini}}, \bibinfo {author} {\bibfnamefont
  {H.}~\bibnamefont {Enriquez}}, \bibinfo {author} {\bibfnamefont
  {S.}~\bibnamefont {Chiang}}, \bibinfo {author} {\bibfnamefont
  {P.}~\bibnamefont {Soukiassian}}, \bibinfo {author} {\bibfnamefont
  {C.}~\bibnamefont {Berger}}, \bibinfo {author} {\bibfnamefont {W.~A.}\
  \bibnamefont {de~Heer}}, \bibinfo {author} {\bibfnamefont {A.}~\bibnamefont
  {Lanzara}}, \ and\ \bibinfo {author} {\bibfnamefont {E.~H.}\ \bibnamefont
  {Conrad}},\ }\href {\doibase 10.1103/PhysRevLett.103.226803} {\bibfield
  {journal} {\bibinfo  {journal} {Phys. Rev. Lett.}\ }\textbf {\bibinfo
  {volume} {103}},\ \bibinfo {pages} {226803} (\bibinfo {year}
  {2009})}\BibitemShut {NoStop}%
\bibitem [{\citenamefont {de~Heer}\ \emph {et~al.}(2010)\citenamefont
  {de~Heer}, \citenamefont {Berger}, \citenamefont {Wu}, \citenamefont
  {Sprinkle}, \citenamefont {Hu}, \citenamefont {Ruan}, \citenamefont
  {Stroscio}, \citenamefont {First}, \citenamefont {Haddon}, \citenamefont
  {Piot}, \citenamefont {Faugeras}, \citenamefont {Potemski},\ and\
  \citenamefont {Moon}}]{tBG_15plus5}%
  \BibitemOpen
  \bibfield  {author} {\bibinfo {author} {\bibfnamefont {W.~A.}\ \bibnamefont
  {de~Heer}}, \bibinfo {author} {\bibfnamefont {C.}~\bibnamefont {Berger}},
  \bibinfo {author} {\bibfnamefont {X.}~\bibnamefont {Wu}}, \bibinfo {author}
  {\bibfnamefont {M.}~\bibnamefont {Sprinkle}}, \bibinfo {author}
  {\bibfnamefont {Y.}~\bibnamefont {Hu}}, \bibinfo {author} {\bibfnamefont
  {M.}~\bibnamefont {Ruan}}, \bibinfo {author} {\bibfnamefont {J.~A.}\
  \bibnamefont {Stroscio}}, \bibinfo {author} {\bibfnamefont {P.~N.}\
  \bibnamefont {First}}, \bibinfo {author} {\bibfnamefont {R.}~\bibnamefont
  {Haddon}}, \bibinfo {author} {\bibfnamefont {B.}~\bibnamefont {Piot}},
  \bibinfo {author} {\bibfnamefont {C.}~\bibnamefont {Faugeras}}, \bibinfo
  {author} {\bibfnamefont {M.}~\bibnamefont {Potemski}}, \ and\ \bibinfo
  {author} {\bibfnamefont {J.-S.}\ \bibnamefont {Moon}},\ }\href {\doibase
  10.1088/0022-3727/43/37/374007} {\bibfield  {journal} {\bibinfo  {journal}
  {Journal of Physics D: Applied Physics}\ }\textbf {\bibinfo {volume} {43}},\
  \bibinfo {pages} {374007} (\bibinfo {year} {2010})}\BibitemShut {NoStop}%
\bibitem [{\citenamefont {Sprinkle}\ \emph {et~al.}(2010)\citenamefont
  {Sprinkle}, \citenamefont {Hicks}, \citenamefont {Tejeda}, \citenamefont
  {Taleb-Ibrahimi}, \citenamefont {F{\`{e}}vre}, \citenamefont {Bertran},
  \citenamefont {Tinkey}, \citenamefont {Clark}, \citenamefont {Soukiassian},
  \citenamefont {Martinotti}, \citenamefont {Hass},\ and\ \citenamefont
  {Conrad}}]{tBG_15plus6}%
  \BibitemOpen
  \bibfield  {author} {\bibinfo {author} {\bibfnamefont {M.}~\bibnamefont
  {Sprinkle}}, \bibinfo {author} {\bibfnamefont {J.}~\bibnamefont {Hicks}},
  \bibinfo {author} {\bibfnamefont {A.}~\bibnamefont {Tejeda}}, \bibinfo
  {author} {\bibfnamefont {A.}~\bibnamefont {Taleb-Ibrahimi}}, \bibinfo
  {author} {\bibfnamefont {P.~L.}\ \bibnamefont {F{\`{e}}vre}}, \bibinfo
  {author} {\bibfnamefont {F.}~\bibnamefont {Bertran}}, \bibinfo {author}
  {\bibfnamefont {H.}~\bibnamefont {Tinkey}}, \bibinfo {author} {\bibfnamefont
  {M.~C.}\ \bibnamefont {Clark}}, \bibinfo {author} {\bibfnamefont
  {P.}~\bibnamefont {Soukiassian}}, \bibinfo {author} {\bibfnamefont
  {D.}~\bibnamefont {Martinotti}}, \bibinfo {author} {\bibfnamefont
  {J.}~\bibnamefont {Hass}}, \ and\ \bibinfo {author} {\bibfnamefont {E.~H.}\
  \bibnamefont {Conrad}},\ }\href {\doibase 10.1088/0022-3727/43/37/374006}
  {\bibfield  {journal} {\bibinfo  {journal} {Journal of Physics D: Applied
  Physics}\ }\textbf {\bibinfo {volume} {43}},\ \bibinfo {pages} {374006}
  (\bibinfo {year} {2010})}\BibitemShut {NoStop}%
\bibitem [{\citenamefont {van Wijk}\ \emph {et~al.}(2015)\citenamefont {van
  Wijk}, \citenamefont {Schuring}, \citenamefont {Katsnelson},\ and\
  \citenamefont {Fasolino}}]{tBG_15plus7}%
  \BibitemOpen
  \bibfield  {author} {\bibinfo {author} {\bibfnamefont {M.~M.}\ \bibnamefont
  {van Wijk}}, \bibinfo {author} {\bibfnamefont {A.}~\bibnamefont {Schuring}},
  \bibinfo {author} {\bibfnamefont {M.~I.}\ \bibnamefont {Katsnelson}}, \ and\
  \bibinfo {author} {\bibfnamefont {A.}~\bibnamefont {Fasolino}},\ }\href
  {\doibase 10.1088/2053-1583/2/3/034010} {\bibfield  {journal} {\bibinfo
  {journal} {2D Materials}\ }\textbf {\bibinfo {volume} {2}},\ \bibinfo {pages}
  {034010} (\bibinfo {year} {2015})}\BibitemShut {NoStop}%
\bibitem [{\citenamefont {Su\'arez~Morell}\ \emph
  {et~al.}(2010{\natexlab{a}})\citenamefont {Su\'arez~Morell}, \citenamefont
  {Correa}, \citenamefont {Vargas}, \citenamefont {Pacheco},\ and\
  \citenamefont {Barticevic}}]{vf_theta0}%
  \BibitemOpen
  \bibfield  {author} {\bibinfo {author} {\bibfnamefont {E.}~\bibnamefont
  {Su\'arez~Morell}}, \bibinfo {author} {\bibfnamefont {J.~D.}\ \bibnamefont
  {Correa}}, \bibinfo {author} {\bibfnamefont {P.}~\bibnamefont {Vargas}},
  \bibinfo {author} {\bibfnamefont {M.}~\bibnamefont {Pacheco}}, \ and\
  \bibinfo {author} {\bibfnamefont {Z.}~\bibnamefont {Barticevic}},\ }\href
  {\doibase 10.1103/PhysRevB.82.121407} {\bibfield  {journal} {\bibinfo
  {journal} {Phys. Rev. B}\ }\textbf {\bibinfo {volume} {82}},\ \bibinfo
  {pages} {121407} (\bibinfo {year} {2010}{\natexlab{a}})}\BibitemShut
  {NoStop}%
\bibitem [{\citenamefont {Yin}\ \emph {et~al.}(2015)\citenamefont {Yin},
  \citenamefont {Qiao}, \citenamefont {Wang}, \citenamefont {Zuo},
  \citenamefont {Yan}, \citenamefont {Xu}, \citenamefont {Dou}, \citenamefont
  {Nie},\ and\ \citenamefont {He}}]{vf_theta1}%
  \BibitemOpen
  \bibfield  {author} {\bibinfo {author} {\bibfnamefont {L.-J.}\ \bibnamefont
  {Yin}}, \bibinfo {author} {\bibfnamefont {J.-B.}\ \bibnamefont {Qiao}},
  \bibinfo {author} {\bibfnamefont {W.-X.}\ \bibnamefont {Wang}}, \bibinfo
  {author} {\bibfnamefont {W.-J.}\ \bibnamefont {Zuo}}, \bibinfo {author}
  {\bibfnamefont {W.}~\bibnamefont {Yan}}, \bibinfo {author} {\bibfnamefont
  {R.}~\bibnamefont {Xu}}, \bibinfo {author} {\bibfnamefont {R.-F.}\
  \bibnamefont {Dou}}, \bibinfo {author} {\bibfnamefont {J.-C.}\ \bibnamefont
  {Nie}}, \ and\ \bibinfo {author} {\bibfnamefont {L.}~\bibnamefont {He}},\
  }\href {\doibase 10.1103/PhysRevB.92.201408} {\bibfield  {journal} {\bibinfo
  {journal} {Phys. Rev. B}\ }\textbf {\bibinfo {volume} {92}},\ \bibinfo
  {pages} {201408} (\bibinfo {year} {2015})}\BibitemShut {NoStop}%
\bibitem [{\citenamefont {Su\'arez~Morell}\ \emph
  {et~al.}(2010{\natexlab{b}})\citenamefont {Su\'arez~Morell}, \citenamefont
  {Correa}, \citenamefont {Vargas}, \citenamefont {Pacheco},\ and\
  \citenamefont {Barticevic}}]{TBG_flat_band}%
  \BibitemOpen
  \bibfield  {author} {\bibinfo {author} {\bibfnamefont {E.}~\bibnamefont
  {Su\'arez~Morell}}, \bibinfo {author} {\bibfnamefont {J.~D.}\ \bibnamefont
  {Correa}}, \bibinfo {author} {\bibfnamefont {P.}~\bibnamefont {Vargas}},
  \bibinfo {author} {\bibfnamefont {M.}~\bibnamefont {Pacheco}}, \ and\
  \bibinfo {author} {\bibfnamefont {Z.}~\bibnamefont {Barticevic}},\ }\href
  {\doibase 10.1103/PhysRevB.82.121407} {\bibfield  {journal} {\bibinfo
  {journal} {Phys. Rev. B}\ }\textbf {\bibinfo {volume} {82}},\ \bibinfo
  {pages} {121407} (\bibinfo {year} {2010}{\natexlab{b}})}\BibitemShut
  {NoStop}%
\bibitem [{\citenamefont {Bistritzer}\ and\ \citenamefont
  {MacDonald}(2011)}]{TBG_flatband1}%
  \BibitemOpen
  \bibfield  {author} {\bibinfo {author} {\bibfnamefont {R.}~\bibnamefont
  {Bistritzer}}\ and\ \bibinfo {author} {\bibfnamefont {A.~H.}\ \bibnamefont
  {MacDonald}},\ }\href {\doibase 10.1073/pnas.1108174108} {\bibfield
  {journal} {\bibinfo  {journal} {Proceedings of the National Academy of
  Sciences}\ }\textbf {\bibinfo {volume} {108}},\ \bibinfo {pages} {12233}
  (\bibinfo {year} {2011})}\BibitemShut {NoStop}%
\bibitem [{\citenamefont {Cao}\ \emph {et~al.}(2018{\natexlab{a}})\citenamefont
  {Cao}, \citenamefont {Fatemi}, \citenamefont {Fang}, \citenamefont
  {Watanabe}, \citenamefont {Taniguchi}, \citenamefont {Kaxiras},\ and\
  \citenamefont {Jarillo-Herrero}}]{BG_superconducting}%
  \BibitemOpen
  \bibfield  {author} {\bibinfo {author} {\bibfnamefont {Y.}~\bibnamefont
  {Cao}}, \bibinfo {author} {\bibfnamefont {V.}~\bibnamefont {Fatemi}},
  \bibinfo {author} {\bibfnamefont {S.}~\bibnamefont {Fang}}, \bibinfo {author}
  {\bibfnamefont {K.}~\bibnamefont {Watanabe}}, \bibinfo {author}
  {\bibfnamefont {T.}~\bibnamefont {Taniguchi}}, \bibinfo {author}
  {\bibfnamefont {E.}~\bibnamefont {Kaxiras}}, \ and\ \bibinfo {author}
  {\bibfnamefont {P.}~\bibnamefont {Jarillo-Herrero}},\ }\href
  {https://doi.org/10.1038/nature26160} {\bibfield  {journal} {\bibinfo
  {journal} {Nature}\ }\textbf {\bibinfo {volume} {556}},\ \bibinfo {pages}
  {43} (\bibinfo {year} {2018}{\natexlab{a}})},\ \bibinfo {note}
  {article}\BibitemShut {NoStop}%
\bibitem [{\citenamefont {Po}\ \emph {et~al.}(2018)\citenamefont {Po},
  \citenamefont {Zou}, \citenamefont {Vishwanath},\ and\ \citenamefont
  {Senthil}}]{TBG_supertivity1}%
  \BibitemOpen
  \bibfield  {author} {\bibinfo {author} {\bibfnamefont {H.~C.}\ \bibnamefont
  {Po}}, \bibinfo {author} {\bibfnamefont {L.}~\bibnamefont {Zou}}, \bibinfo
  {author} {\bibfnamefont {A.}~\bibnamefont {Vishwanath}}, \ and\ \bibinfo
  {author} {\bibfnamefont {T.}~\bibnamefont {Senthil}},\ }\href {\doibase
  10.1103/PhysRevX.8.031089} {\bibfield  {journal} {\bibinfo  {journal} {Phys.
  Rev. X}\ }\textbf {\bibinfo {volume} {8}},\ \bibinfo {pages} {031089}
  (\bibinfo {year} {2018})}\BibitemShut {NoStop}%
\bibitem [{\citenamefont {Yankowitz}\ \emph {et~al.}(2019)\citenamefont
  {Yankowitz}, \citenamefont {Chen}, \citenamefont {Polshyn}, \citenamefont
  {Zhang}, \citenamefont {Watanabe}, \citenamefont {Taniguchi}, \citenamefont
  {Graf}, \citenamefont {Young},\ and\ \citenamefont
  {Dean}}]{TBG_supertivity2}%
  \BibitemOpen
  \bibfield  {author} {\bibinfo {author} {\bibfnamefont {M.}~\bibnamefont
  {Yankowitz}}, \bibinfo {author} {\bibfnamefont {S.}~\bibnamefont {Chen}},
  \bibinfo {author} {\bibfnamefont {H.}~\bibnamefont {Polshyn}}, \bibinfo
  {author} {\bibfnamefont {Y.}~\bibnamefont {Zhang}}, \bibinfo {author}
  {\bibfnamefont {K.}~\bibnamefont {Watanabe}}, \bibinfo {author}
  {\bibfnamefont {T.}~\bibnamefont {Taniguchi}}, \bibinfo {author}
  {\bibfnamefont {D.}~\bibnamefont {Graf}}, \bibinfo {author} {\bibfnamefont
  {A.~F.}\ \bibnamefont {Young}}, \ and\ \bibinfo {author} {\bibfnamefont
  {C.~R.}\ \bibnamefont {Dean}},\ }\href {\doibase 10.1126/science.aav1910}
  {\bibfield  {journal} {\bibinfo  {journal} {Science}\ }\textbf {\bibinfo
  {volume} {363}},\ \bibinfo {pages} {1059} (\bibinfo {year}
  {2019})}\BibitemShut {NoStop}%
\bibitem [{\citenamefont {Cao}\ \emph {et~al.}(2018{\natexlab{b}})\citenamefont
  {Cao}, \citenamefont {Fatemi}, \citenamefont {Demir}, \citenamefont {Fang},
  \citenamefont {Tomarken}, \citenamefont {Luo}, \citenamefont
  {Sanchez-Yamagishi}, \citenamefont {Watanabe}, \citenamefont {Taniguchi},
  \citenamefont {Kaxiras}, \citenamefont {Ashoori},\ and\ \citenamefont
  {Jarillo-Herrero}}]{TBG_insulator_phase}%
  \BibitemOpen
  \bibfield  {author} {\bibinfo {author} {\bibfnamefont {Y.}~\bibnamefont
  {Cao}}, \bibinfo {author} {\bibfnamefont {V.}~\bibnamefont {Fatemi}},
  \bibinfo {author} {\bibfnamefont {A.}~\bibnamefont {Demir}}, \bibinfo
  {author} {\bibfnamefont {S.}~\bibnamefont {Fang}}, \bibinfo {author}
  {\bibfnamefont {S.~L.}\ \bibnamefont {Tomarken}}, \bibinfo {author}
  {\bibfnamefont {J.~Y.}\ \bibnamefont {Luo}}, \bibinfo {author} {\bibfnamefont
  {J.~D.}\ \bibnamefont {Sanchez-Yamagishi}}, \bibinfo {author} {\bibfnamefont
  {K.}~\bibnamefont {Watanabe}}, \bibinfo {author} {\bibfnamefont
  {T.}~\bibnamefont {Taniguchi}}, \bibinfo {author} {\bibfnamefont
  {E.}~\bibnamefont {Kaxiras}}, \bibinfo {author} {\bibfnamefont {R.~C.}\
  \bibnamefont {Ashoori}}, \ and\ \bibinfo {author} {\bibfnamefont
  {P.}~\bibnamefont {Jarillo-Herrero}},\ }\href
  {https://doi.org/10.1038/nature26154} {\bibfield  {journal} {\bibinfo
  {journal} {Nature}\ }\textbf {\bibinfo {volume} {556}},\ \bibinfo {pages}
  {80} (\bibinfo {year} {2018}{\natexlab{b}})}\BibitemShut {NoStop}%
\bibitem [{\citenamefont {Ahn}\ \emph {et~al.}(2018)\citenamefont {Ahn},
  \citenamefont {Moon}, \citenamefont {Kim}, \citenamefont {Kim}, \citenamefont
  {Shin}, \citenamefont {Kim}, \citenamefont {Cha}, \citenamefont {Kahng},
  \citenamefont {Kim}, \citenamefont {Koshino}, \citenamefont {Son},
  \citenamefont {Yang},\ and\ \citenamefont {Ahn}}]{science_QC}%
  \BibitemOpen
  \bibfield  {author} {\bibinfo {author} {\bibfnamefont {S.~J.}\ \bibnamefont
  {Ahn}}, \bibinfo {author} {\bibfnamefont {P.}~\bibnamefont {Moon}}, \bibinfo
  {author} {\bibfnamefont {T.-H.}\ \bibnamefont {Kim}}, \bibinfo {author}
  {\bibfnamefont {H.-W.}\ \bibnamefont {Kim}}, \bibinfo {author} {\bibfnamefont
  {H.-C.}\ \bibnamefont {Shin}}, \bibinfo {author} {\bibfnamefont {E.~H.}\
  \bibnamefont {Kim}}, \bibinfo {author} {\bibfnamefont {H.~W.}\ \bibnamefont
  {Cha}}, \bibinfo {author} {\bibfnamefont {S.-J.}\ \bibnamefont {Kahng}},
  \bibinfo {author} {\bibfnamefont {P.}~\bibnamefont {Kim}}, \bibinfo {author}
  {\bibfnamefont {M.}~\bibnamefont {Koshino}}, \bibinfo {author} {\bibfnamefont
  {Y.-W.}\ \bibnamefont {Son}}, \bibinfo {author} {\bibfnamefont {C.-W.}\
  \bibnamefont {Yang}}, \ and\ \bibinfo {author} {\bibfnamefont {J.~R.}\
  \bibnamefont {Ahn}},\ }\href {\doibase 10.1126/science.aar8412} {\bibfield
  {journal} {\bibinfo  {journal} {Science}\ }\textbf {\bibinfo {volume}
  {361}},\ \bibinfo {pages} {782} (\bibinfo {year} {2018})}\BibitemShut
  {NoStop}%
\bibitem [{\citenamefont {Yao}\ \emph {et~al.}(2018)\citenamefont {Yao},
  \citenamefont {Wang}, \citenamefont {Bao}, \citenamefont {Zhang},
  \citenamefont {Zhang}, \citenamefont {Bao}, \citenamefont {Chan},
  \citenamefont {Chen}, \citenamefont {Avila}, \citenamefont {Asensio},
  \citenamefont {Zhu},\ and\ \citenamefont {Zhou}}]{pnas_QC}%
  \BibitemOpen
  \bibfield  {author} {\bibinfo {author} {\bibfnamefont {W.}~\bibnamefont
  {Yao}}, \bibinfo {author} {\bibfnamefont {E.}~\bibnamefont {Wang}}, \bibinfo
  {author} {\bibfnamefont {C.}~\bibnamefont {Bao}}, \bibinfo {author}
  {\bibfnamefont {Y.}~\bibnamefont {Zhang}}, \bibinfo {author} {\bibfnamefont
  {K.}~\bibnamefont {Zhang}}, \bibinfo {author} {\bibfnamefont
  {K.}~\bibnamefont {Bao}}, \bibinfo {author} {\bibfnamefont {C.~K.}\
  \bibnamefont {Chan}}, \bibinfo {author} {\bibfnamefont {C.}~\bibnamefont
  {Chen}}, \bibinfo {author} {\bibfnamefont {J.}~\bibnamefont {Avila}},
  \bibinfo {author} {\bibfnamefont {M.~C.}\ \bibnamefont {Asensio}}, \bibinfo
  {author} {\bibfnamefont {J.}~\bibnamefont {Zhu}}, \ and\ \bibinfo {author}
  {\bibfnamefont {S.}~\bibnamefont {Zhou}},\ }\href {\doibase
  10.1073/pnas.1720865115} {\bibfield  {journal} {\bibinfo  {journal}
  {Proceedings of the National Academy of Sciences}\ }\textbf {\bibinfo
  {volume} {115}},\ \bibinfo {pages} {6928} (\bibinfo {year}
  {2018})}\BibitemShut {NoStop}%
\bibitem [{\citenamefont {Takesaki}\ \emph {et~al.}(2016)\citenamefont
  {Takesaki}, \citenamefont {Kawahara}, \citenamefont {Hibino}, \citenamefont
  {Okada}, \citenamefont {Tsuji},\ and\ \citenamefont {Ago}}]{cm_QC}%
  \BibitemOpen
  \bibfield  {author} {\bibinfo {author} {\bibfnamefont {Y.}~\bibnamefont
  {Takesaki}}, \bibinfo {author} {\bibfnamefont {K.}~\bibnamefont {Kawahara}},
  \bibinfo {author} {\bibfnamefont {H.}~\bibnamefont {Hibino}}, \bibinfo
  {author} {\bibfnamefont {S.}~\bibnamefont {Okada}}, \bibinfo {author}
  {\bibfnamefont {M.}~\bibnamefont {Tsuji}}, \ and\ \bibinfo {author}
  {\bibfnamefont {H.}~\bibnamefont {Ago}},\ }\href {\doibase
  10.1021/acs.chemmater.6b01137} {\bibfield  {journal} {\bibinfo  {journal}
  {Chemistry of Materials}\ }\textbf {\bibinfo {volume} {28}},\ \bibinfo
  {pages} {4583} (\bibinfo {year} {2016})}\BibitemShut {NoStop}%
\bibitem [{\citenamefont {Pezzini}\ \emph {et~al.}(0)\citenamefont {Pezzini},
  \citenamefont {Mišeikis}, \citenamefont {Piccinini}, \citenamefont {Forti},
  \citenamefont {Pace}, \citenamefont {Engelke}, \citenamefont {Rossella},
  \citenamefont {Watanabe}, \citenamefont {Taniguchi}, \citenamefont {Kim},\
  and\ \citenamefont {Coletti}}]{30tBG_onCu_arXiv}%
  \BibitemOpen
  \bibfield  {author} {\bibinfo {author} {\bibfnamefont {S.}~\bibnamefont
  {Pezzini}}, \bibinfo {author} {\bibfnamefont {V.}~\bibnamefont {Mišeikis}},
  \bibinfo {author} {\bibfnamefont {G.}~\bibnamefont {Piccinini}}, \bibinfo
  {author} {\bibfnamefont {S.}~\bibnamefont {Forti}}, \bibinfo {author}
  {\bibfnamefont {S.}~\bibnamefont {Pace}}, \bibinfo {author} {\bibfnamefont
  {R.}~\bibnamefont {Engelke}}, \bibinfo {author} {\bibfnamefont
  {F.}~\bibnamefont {Rossella}}, \bibinfo {author} {\bibfnamefont
  {K.}~\bibnamefont {Watanabe}}, \bibinfo {author} {\bibfnamefont
  {T.}~\bibnamefont {Taniguchi}}, \bibinfo {author} {\bibfnamefont
  {P.}~\bibnamefont {Kim}}, \ and\ \bibinfo {author} {\bibfnamefont
  {C.}~\bibnamefont {Coletti}},\ }\href {\doibase 10.1021/acs.nanolett.0c00172}
  {\bibfield  {journal} {\bibinfo  {journal} {Nano Letters}\ }\textbf {\bibinfo
  {volume} {0}},\ \bibinfo {pages} {null} (\bibinfo {year} {0})},\ \bibinfo
  {note} {pMID: 32297749},\ \Eprint
  {http://arxiv.org/abs/https://doi.org/10.1021/acs.nanolett.0c00172}
  {https://doi.org/10.1021/acs.nanolett.0c00172} \BibitemShut {NoStop}%
\bibitem [{\citenamefont {Deng}\ \emph
  {et~al.}(2020{\natexlab{a}})\citenamefont {Deng}, \citenamefont {Wang},
  \citenamefont {Li}, \citenamefont {Li}, \citenamefont {Wang}, \citenamefont
  {Tang}, \citenamefont {Fu}, \citenamefont {Tian}, \citenamefont {Gao},
  \citenamefont {Xue},\ and\ \citenamefont {Peng}}]{30tBG_on_Cu_ACSNano}%
  \BibitemOpen
  \bibfield  {author} {\bibinfo {author} {\bibfnamefont {B.}~\bibnamefont
  {Deng}}, \bibinfo {author} {\bibfnamefont {B.}~\bibnamefont {Wang}}, \bibinfo
  {author} {\bibfnamefont {N.}~\bibnamefont {Li}}, \bibinfo {author}
  {\bibfnamefont {R.}~\bibnamefont {Li}}, \bibinfo {author} {\bibfnamefont
  {Y.}~\bibnamefont {Wang}}, \bibinfo {author} {\bibfnamefont {J.}~\bibnamefont
  {Tang}}, \bibinfo {author} {\bibfnamefont {Q.}~\bibnamefont {Fu}}, \bibinfo
  {author} {\bibfnamefont {Z.}~\bibnamefont {Tian}}, \bibinfo {author}
  {\bibfnamefont {P.}~\bibnamefont {Gao}}, \bibinfo {author} {\bibfnamefont
  {J.}~\bibnamefont {Xue}}, \ and\ \bibinfo {author} {\bibfnamefont
  {H.}~\bibnamefont {Peng}},\ }\href {\doibase 10.1021/acsnano.9b07091}
  {\bibfield  {journal} {\bibinfo  {journal} {ACS Nano}\ }\textbf {\bibinfo
  {volume} {14}},\ \bibinfo {pages} {1656} (\bibinfo {year}
  {2020}{\natexlab{a}})}\BibitemShut {NoStop}%
\bibitem [{\citenamefont {{Lin}}\ \emph {et~al.}(2018)\citenamefont {{Lin}},
  \citenamefont {{Samiseresht}}, \citenamefont {{Franke}}, \citenamefont
  {{Parhizkar}}, \citenamefont {{Soubatch}}, \citenamefont {{Amorim}},
  \citenamefont {{Lee}}, \citenamefont {{Kumpf}}, \citenamefont {{Tautz}},\
  and\ \citenamefont {{Bocquet}}}]{30TBG_grown}%
  \BibitemOpen
  \bibfield  {author} {\bibinfo {author} {\bibfnamefont {Y.~R.}\ \bibnamefont
  {{Lin}}}, \bibinfo {author} {\bibfnamefont {N.}~\bibnamefont
  {{Samiseresht}}}, \bibinfo {author} {\bibfnamefont {M.}~\bibnamefont
  {{Franke}}}, \bibinfo {author} {\bibfnamefont {S.}~\bibnamefont
  {{Parhizkar}}}, \bibinfo {author} {\bibfnamefont {S.}~\bibnamefont
  {{Soubatch}}}, \bibinfo {author} {\bibfnamefont {B.}~\bibnamefont
  {{Amorim}}}, \bibinfo {author} {\bibfnamefont {T.~L.}\ \bibnamefont {{Lee}}},
  \bibinfo {author} {\bibfnamefont {C.}~\bibnamefont {{Kumpf}}}, \bibinfo
  {author} {\bibfnamefont {F.~S.}\ \bibnamefont {{Tautz}}}, \ and\ \bibinfo
  {author} {\bibfnamefont {F.~C.}\ \bibnamefont {{Bocquet}}},\ }\href
  {https://arxiv.org/abs/1809.07958} {\bibfield  {journal} {\bibinfo  {journal}
  {arXiv e-prints}\ ,\ \bibinfo {pages} {arXiv:1809.07958}} (\bibinfo {year}
  {2018})}\BibitemShut {NoStop}%
\bibitem [{\citenamefont {Park}\ \emph {et~al.}(2019)\citenamefont {Park},
  \citenamefont {Kim},\ and\ \citenamefont {Lee}}]{30TBG_localization}%
  \BibitemOpen
  \bibfield  {author} {\bibinfo {author} {\bibfnamefont {M.~J.}\ \bibnamefont
  {Park}}, \bibinfo {author} {\bibfnamefont {H.~S.}\ \bibnamefont {Kim}}, \
  and\ \bibinfo {author} {\bibfnamefont {S.}~\bibnamefont {Lee}},\ }\href
  {\doibase 10.1103/PhysRevB.99.245401} {\bibfield  {journal} {\bibinfo
  {journal} {Phys. Rev. B}\ }\textbf {\bibinfo {volume} {99}},\ \bibinfo
  {pages} {245401} (\bibinfo {year} {2019})}\BibitemShut {NoStop}%
\bibitem [{\citenamefont {Koren}\ and\ \citenamefont
  {Duerig}(2016)}]{30TBG_superlubricity}%
  \BibitemOpen
  \bibfield  {author} {\bibinfo {author} {\bibfnamefont {E.}~\bibnamefont
  {Koren}}\ and\ \bibinfo {author} {\bibfnamefont {U.}~\bibnamefont {Duerig}},\
  }\href {\doibase 10.1103/PhysRevB.93.201404} {\bibfield  {journal} {\bibinfo
  {journal} {Phys. Rev. B}\ }\textbf {\bibinfo {volume} {93}},\ \bibinfo
  {pages} {201404} (\bibinfo {year} {2016})}\BibitemShut {NoStop}%
\bibitem [{\citenamefont {Suzuki}\ \emph {et~al.}(2019)\citenamefont {Suzuki},
  \citenamefont {Iimori}, \citenamefont {Ahn}, \citenamefont {Zhao},
  \citenamefont {Watanabe}, \citenamefont {Xu}, \citenamefont {Fujisawa},
  \citenamefont {Kanai}, \citenamefont {Ishii}, \citenamefont {Itatani},
  \citenamefont {Suwa}, \citenamefont {Fukidome}, \citenamefont {Tanaka},
  \citenamefont {Ahn}, \citenamefont {Okazaki}, \citenamefont {Shin},
  \citenamefont {Komori},\ and\ \citenamefont {Matsuda}}]{30TBG_Suzuki}%
  \BibitemOpen
  \bibfield  {author} {\bibinfo {author} {\bibfnamefont {T.}~\bibnamefont
  {Suzuki}}, \bibinfo {author} {\bibfnamefont {T.}~\bibnamefont {Iimori}},
  \bibinfo {author} {\bibfnamefont {S.~J.}\ \bibnamefont {Ahn}}, \bibinfo
  {author} {\bibfnamefont {Y.}~\bibnamefont {Zhao}}, \bibinfo {author}
  {\bibfnamefont {M.}~\bibnamefont {Watanabe}}, \bibinfo {author}
  {\bibfnamefont {J.}~\bibnamefont {Xu}}, \bibinfo {author} {\bibfnamefont
  {M.}~\bibnamefont {Fujisawa}}, \bibinfo {author} {\bibfnamefont
  {T.}~\bibnamefont {Kanai}}, \bibinfo {author} {\bibfnamefont
  {N.}~\bibnamefont {Ishii}}, \bibinfo {author} {\bibfnamefont
  {J.}~\bibnamefont {Itatani}}, \bibinfo {author} {\bibfnamefont
  {K.}~\bibnamefont {Suwa}}, \bibinfo {author} {\bibfnamefont {H.}~\bibnamefont
  {Fukidome}}, \bibinfo {author} {\bibfnamefont {S.}~\bibnamefont {Tanaka}},
  \bibinfo {author} {\bibfnamefont {J.~R.}\ \bibnamefont {Ahn}}, \bibinfo
  {author} {\bibfnamefont {K.}~\bibnamefont {Okazaki}}, \bibinfo {author}
  {\bibfnamefont {S.}~\bibnamefont {Shin}}, \bibinfo {author} {\bibfnamefont
  {F.}~\bibnamefont {Komori}}, \ and\ \bibinfo {author} {\bibfnamefont
  {I.}~\bibnamefont {Matsuda}},\ }\href {\doibase 10.1021/acsnano.9b06091}
  {\bibfield  {journal} {\bibinfo  {journal} {ACS Nano}\ }\textbf {\bibinfo
  {volume} {13}},\ \bibinfo {pages} {11981} (\bibinfo {year}
  {2019})}\BibitemShut {NoStop}%
\bibitem [{\citenamefont {Moon}\ \emph {et~al.}(2019)\citenamefont {Moon},
  \citenamefont {Koshino},\ and\ \citenamefont {Son}}]{arXiv_QC}%
  \BibitemOpen
  \bibfield  {author} {\bibinfo {author} {\bibfnamefont {P.}~\bibnamefont
  {Moon}}, \bibinfo {author} {\bibfnamefont {M.}~\bibnamefont {Koshino}}, \
  and\ \bibinfo {author} {\bibfnamefont {Y.-W.}\ \bibnamefont {Son}},\ }\href
  {\doibase 10.1103/PhysRevB.99.165430} {\bibfield  {journal} {\bibinfo
  {journal} {Phys. Rev. B}\ }\textbf {\bibinfo {volume} {99}},\ \bibinfo
  {pages} {165430} (\bibinfo {year} {2019})}\BibitemShut {NoStop}%
\bibitem [{\citenamefont {Yan}\ \emph {et~al.}(2019{\natexlab{a}})\citenamefont
  {Yan}, \citenamefont {Ma}, \citenamefont {Qiao}, \citenamefont {Zhong},
  \citenamefont {Yang}, \citenamefont {Li}, \citenamefont {Fu}, \citenamefont
  {Zhang},\ and\ \citenamefont {He}}]{TBG_2dM}%
  \BibitemOpen
  \bibfield  {author} {\bibinfo {author} {\bibfnamefont {C.}~\bibnamefont
  {Yan}}, \bibinfo {author} {\bibfnamefont {D.}~\bibnamefont {Ma}}, \bibinfo
  {author} {\bibfnamefont {J.}~\bibnamefont {Qiao}}, \bibinfo {author}
  {\bibfnamefont {H.}~\bibnamefont {Zhong}}, \bibinfo {author} {\bibfnamefont
  {L.}~\bibnamefont {Yang}}, \bibinfo {author} {\bibfnamefont {S.-Y.}\
  \bibnamefont {Li}}, \bibinfo {author} {\bibfnamefont {Z.}~\bibnamefont {Fu}},
  \bibinfo {author} {\bibfnamefont {Y.}~\bibnamefont {Zhang}}, \ and\ \bibinfo
  {author} {\bibfnamefont {L.}~\bibnamefont {He}},\ }\href
  {http://iopscience.iop.org/10.1088/2053-1583/ab3b16} {\bibfield  {journal}
  {\bibinfo  {journal} {2D Materials}\ } (\bibinfo {year}
  {2019}{\natexlab{a}})}\BibitemShut {NoStop}%
\bibitem [{\citenamefont {Yu}\ \emph {et~al.}(2019)\citenamefont {Yu},
  \citenamefont {Wu}, \citenamefont {Zhan}, \citenamefont {Katsnelson},\ and\
  \citenamefont {Yuan}}]{Yu_QC_BG}%
  \BibitemOpen
  \bibfield  {author} {\bibinfo {author} {\bibfnamefont {G.}~\bibnamefont
  {Yu}}, \bibinfo {author} {\bibfnamefont {Z.}~\bibnamefont {Wu}}, \bibinfo
  {author} {\bibfnamefont {Z.}~\bibnamefont {Zhan}}, \bibinfo {author}
  {\bibfnamefont {M.~I.}\ \bibnamefont {Katsnelson}}, \ and\ \bibinfo {author}
  {\bibfnamefont {S.}~\bibnamefont {Yuan}},\ }\href {\doibase
  10.1038/s41524-019-0258-0} {\bibfield  {journal} {\bibinfo  {journal} {npj
  Computational Materials}\ }\textbf {\bibinfo {volume} {5}},\ \bibinfo {pages}
  {122} (\bibinfo {year} {2019})}\BibitemShut {NoStop}%
\bibitem [{\citenamefont {Yan}\ \emph {et~al.}(2019{\natexlab{b}})\citenamefont
  {Yan}, \citenamefont {Ma}, \citenamefont {Qiao}, \citenamefont {Zhong},
  \citenamefont {Yang}, \citenamefont {Li}, \citenamefont {Fu}, \citenamefont
  {Zhang},\ and\ \citenamefont {He}}]{30TBG_STM}%
  \BibitemOpen
  \bibfield  {author} {\bibinfo {author} {\bibfnamefont {C.}~\bibnamefont
  {Yan}}, \bibinfo {author} {\bibfnamefont {D.-L.}\ \bibnamefont {Ma}},
  \bibinfo {author} {\bibfnamefont {J.-B.}\ \bibnamefont {Qiao}}, \bibinfo
  {author} {\bibfnamefont {H.-Y.}\ \bibnamefont {Zhong}}, \bibinfo {author}
  {\bibfnamefont {L.}~\bibnamefont {Yang}}, \bibinfo {author} {\bibfnamefont
  {S.-Y.}\ \bibnamefont {Li}}, \bibinfo {author} {\bibfnamefont {Z.-Q.}\
  \bibnamefont {Fu}}, \bibinfo {author} {\bibfnamefont {Y.}~\bibnamefont
  {Zhang}}, \ and\ \bibinfo {author} {\bibfnamefont {L.}~\bibnamefont {He}},\
  }\href {\doibase 10.1088/2053-1583/ab3b16} {\bibfield  {journal} {\bibinfo
  {journal} {2D Materials}\ }\textbf {\bibinfo {volume} {6}},\ \bibinfo {pages}
  {045041} (\bibinfo {year} {2019}{\natexlab{b}})}\BibitemShut {NoStop}%
\bibitem [{\citenamefont {Deng}\ \emph
  {et~al.}(2020{\natexlab{b}})\citenamefont {Deng}, \citenamefont {Wang},
  \citenamefont {Li}, \citenamefont {Li}, \citenamefont {Wang}, \citenamefont
  {Tang}, \citenamefont {Fu}, \citenamefont {Tian}, \citenamefont {Gao},
  \citenamefont {Xue},\ and\ \citenamefont {Peng}}]{decoupling_acs}%
  \BibitemOpen
  \bibfield  {author} {\bibinfo {author} {\bibfnamefont {B.}~\bibnamefont
  {Deng}}, \bibinfo {author} {\bibfnamefont {B.}~\bibnamefont {Wang}}, \bibinfo
  {author} {\bibfnamefont {N.}~\bibnamefont {Li}}, \bibinfo {author}
  {\bibfnamefont {R.}~\bibnamefont {Li}}, \bibinfo {author} {\bibfnamefont
  {Y.}~\bibnamefont {Wang}}, \bibinfo {author} {\bibfnamefont {J.}~\bibnamefont
  {Tang}}, \bibinfo {author} {\bibfnamefont {Q.}~\bibnamefont {Fu}}, \bibinfo
  {author} {\bibfnamefont {Z.}~\bibnamefont {Tian}}, \bibinfo {author}
  {\bibfnamefont {P.}~\bibnamefont {Gao}}, \bibinfo {author} {\bibfnamefont
  {J.}~\bibnamefont {Xue}}, \ and\ \bibinfo {author} {\bibfnamefont
  {H.}~\bibnamefont {Peng}},\ }\href {\doibase 10.1021/acsnano.9b07091}
  {\bibfield  {journal} {\bibinfo  {journal} {ACS Nano}\ }\textbf {\bibinfo
  {volume} {14}},\ \bibinfo {pages} {1656} (\bibinfo {year}
  {2020}{\natexlab{b}})}\BibitemShut {NoStop}%
\bibitem [{\citenamefont {{Liu}}\ \emph {et~al.}(2019)\citenamefont {{Liu}},
  \citenamefont {{Hao}}, \citenamefont {{Khalaf}}, \citenamefont {{Lee}},
  \citenamefont {{Watanabe}}, \citenamefont {{Taniguchi}}, \citenamefont
  {{Vishwanath}},\ and\ \citenamefont {{Kim}}}]{TDBG_spin_polarized_supercond}%
  \BibitemOpen
  \bibfield  {author} {\bibinfo {author} {\bibfnamefont {X.}~\bibnamefont
  {{Liu}}}, \bibinfo {author} {\bibfnamefont {Z.}~\bibnamefont {{Hao}}},
  \bibinfo {author} {\bibfnamefont {E.}~\bibnamefont {{Khalaf}}}, \bibinfo
  {author} {\bibfnamefont {J.~Y.}\ \bibnamefont {{Lee}}}, \bibinfo {author}
  {\bibfnamefont {K.}~\bibnamefont {{Watanabe}}}, \bibinfo {author}
  {\bibfnamefont {T.}~\bibnamefont {{Taniguchi}}}, \bibinfo {author}
  {\bibfnamefont {A.}~\bibnamefont {{Vishwanath}}}, \ and\ \bibinfo {author}
  {\bibfnamefont {P.}~\bibnamefont {{Kim}}},\ }\href
  {https://arxiv.org/abs/1903.08130} {\bibfield  {journal} {\bibinfo  {journal}
  {arXiv e-prints}\ ,\ \bibinfo {pages} {arXiv:1903.08130}} (\bibinfo {year}
  {2019})}\BibitemShut {NoStop}%
\bibitem [{\citenamefont {{Cao}}\ \emph {et~al.}(2019)\citenamefont {{Cao}},
  \citenamefont {{Rodan-Legrain}}, \citenamefont {{Rubies-Bigord{\`a}}},
  \citenamefont {{Park}}, \citenamefont {{Watanabe}}, \citenamefont
  {{Taniguchi}},\ and\ \citenamefont {{Jarillo-Herrero}}}]{TDBG1}%
  \BibitemOpen
  \bibfield  {author} {\bibinfo {author} {\bibfnamefont {Y.}~\bibnamefont
  {{Cao}}}, \bibinfo {author} {\bibfnamefont {D.}~\bibnamefont
  {{Rodan-Legrain}}}, \bibinfo {author} {\bibfnamefont {O.}~\bibnamefont
  {{Rubies-Bigord{\`a}}}}, \bibinfo {author} {\bibfnamefont {J.~M.}\
  \bibnamefont {{Park}}}, \bibinfo {author} {\bibfnamefont {K.}~\bibnamefont
  {{Watanabe}}}, \bibinfo {author} {\bibfnamefont {T.}~\bibnamefont
  {{Taniguchi}}}, \ and\ \bibinfo {author} {\bibfnamefont {P.}~\bibnamefont
  {{Jarillo-Herrero}}},\ }\href {https://arxiv.org/abs/1903.08596} {\bibfield
  {journal} {\bibinfo  {journal} {arXiv e-prints}\ ,\ \bibinfo {pages}
  {arXiv:1903.08596}} (\bibinfo {year} {2019})}\BibitemShut {NoStop}%
\bibitem [{\citenamefont {{Scheurer}}\ \emph {et~al.}(2019)\citenamefont
  {{Scheurer}}, \citenamefont {{Samajdar}},\ and\ \citenamefont
  {{Sachdev}}}]{TDBG2}%
  \BibitemOpen
  \bibfield  {author} {\bibinfo {author} {\bibfnamefont {M.~S.}\ \bibnamefont
  {{Scheurer}}}, \bibinfo {author} {\bibfnamefont {R.}~\bibnamefont
  {{Samajdar}}}, \ and\ \bibinfo {author} {\bibfnamefont {S.}~\bibnamefont
  {{Sachdev}}},\ }\href {https://arxiv.org/abs/1906.03258} {\bibfield
  {journal} {\bibinfo  {journal} {arXiv e-prints}\ ,\ \bibinfo {pages}
  {arXiv:1906.03258}} (\bibinfo {year} {2019})}\BibitemShut {NoStop}%
\bibitem [{\citenamefont {Shen}\ \emph {et~al.}(2020)\citenamefont {Shen},
  \citenamefont {Chu}, \citenamefont {Wu}, \citenamefont {Li}, \citenamefont
  {Wang}, \citenamefont {Zhao}, \citenamefont {Tang}, \citenamefont {Liu},
  \citenamefont {Tian}, \citenamefont {Watanabe}, \citenamefont {Taniguchi},
  \citenamefont {Yang}, \citenamefont {Meng}, \citenamefont {Shi},
  \citenamefont {Yazyev},\ and\ \citenamefont {Zhang}}]{TDBG3}%
  \BibitemOpen
  \bibfield  {author} {\bibinfo {author} {\bibfnamefont {C.}~\bibnamefont
  {Shen}}, \bibinfo {author} {\bibfnamefont {Y.}~\bibnamefont {Chu}}, \bibinfo
  {author} {\bibfnamefont {Q.}~\bibnamefont {Wu}}, \bibinfo {author}
  {\bibfnamefont {N.}~\bibnamefont {Li}}, \bibinfo {author} {\bibfnamefont
  {S.}~\bibnamefont {Wang}}, \bibinfo {author} {\bibfnamefont {Y.}~\bibnamefont
  {Zhao}}, \bibinfo {author} {\bibfnamefont {J.}~\bibnamefont {Tang}}, \bibinfo
  {author} {\bibfnamefont {J.}~\bibnamefont {Liu}}, \bibinfo {author}
  {\bibfnamefont {J.}~\bibnamefont {Tian}}, \bibinfo {author} {\bibfnamefont
  {K.}~\bibnamefont {Watanabe}}, \bibinfo {author} {\bibfnamefont
  {T.}~\bibnamefont {Taniguchi}}, \bibinfo {author} {\bibfnamefont
  {R.}~\bibnamefont {Yang}}, \bibinfo {author} {\bibfnamefont {Z.~Y.}\
  \bibnamefont {Meng}}, \bibinfo {author} {\bibfnamefont {D.}~\bibnamefont
  {Shi}}, \bibinfo {author} {\bibfnamefont {O.~V.}\ \bibnamefont {Yazyev}}, \
  and\ \bibinfo {author} {\bibfnamefont {G.}~\bibnamefont {Zhang}},\ }\href
  {\doibase 10.1038/s41567-020-0825-9} {\bibfield  {journal} {\bibinfo
  {journal} {Nature Physics}\ } (\bibinfo {year} {2020}),\
  10.1038/s41567-020-0825-9}\BibitemShut {NoStop}%
\bibitem [{\citenamefont {Burg}\ \emph {et~al.}(2019)\citenamefont {Burg},
  \citenamefont {Zhu}, \citenamefont {Taniguchi}, \citenamefont {Watanabe},
  \citenamefont {MacDonald},\ and\ \citenamefont {Tutuc}}]{TDBG_insulating}%
  \BibitemOpen
  \bibfield  {author} {\bibinfo {author} {\bibfnamefont {G.~W.}\ \bibnamefont
  {Burg}}, \bibinfo {author} {\bibfnamefont {J.}~\bibnamefont {Zhu}}, \bibinfo
  {author} {\bibfnamefont {T.}~\bibnamefont {Taniguchi}}, \bibinfo {author}
  {\bibfnamefont {K.}~\bibnamefont {Watanabe}}, \bibinfo {author}
  {\bibfnamefont {A.~H.}\ \bibnamefont {MacDonald}}, \ and\ \bibinfo {author}
  {\bibfnamefont {E.}~\bibnamefont {Tutuc}},\ }\href {\doibase
  10.1103/PhysRevLett.123.197702} {\bibfield  {journal} {\bibinfo  {journal}
  {Phys. Rev. Lett.}\ }\textbf {\bibinfo {volume} {123}},\ \bibinfo {pages}
  {197702} (\bibinfo {year} {2019})}\BibitemShut {NoStop}%
\bibitem [{\citenamefont {Haddadi}\ \emph {et~al.}(2020)\citenamefont
  {Haddadi}, \citenamefont {Wu}, \citenamefont {Kruchkov},\ and\ \citenamefont
  {Yazyev}}]{TDBG_flatband0}%
  \BibitemOpen
  \bibfield  {author} {\bibinfo {author} {\bibfnamefont {F.}~\bibnamefont
  {Haddadi}}, \bibinfo {author} {\bibfnamefont {Q.}~\bibnamefont {Wu}},
  \bibinfo {author} {\bibfnamefont {A.~J.}\ \bibnamefont {Kruchkov}}, \ and\
  \bibinfo {author} {\bibfnamefont {O.~V.}\ \bibnamefont {Yazyev}},\ }\href
  {\doibase 10.1021/acs.nanolett.9b05117} {\bibfield  {journal} {\bibinfo
  {journal} {Nano Letters}\ }\textbf {\bibinfo {volume} {20}},\ \bibinfo
  {pages} {2410} (\bibinfo {year} {2020})}\BibitemShut {NoStop}%
\bibitem [{\citenamefont {Chebrolu}\ \emph {et~al.}(2019)\citenamefont
  {Chebrolu}, \citenamefont {Chittari},\ and\ \citenamefont
  {Jung}}]{TDBG_flatband1}%
  \BibitemOpen
  \bibfield  {author} {\bibinfo {author} {\bibfnamefont {N.~R.}\ \bibnamefont
  {Chebrolu}}, \bibinfo {author} {\bibfnamefont {B.~L.}\ \bibnamefont
  {Chittari}}, \ and\ \bibinfo {author} {\bibfnamefont {J.}~\bibnamefont
  {Jung}},\ }\href {\doibase 10.1103/PhysRevB.99.235417} {\bibfield  {journal}
  {\bibinfo  {journal} {Phys. Rev. B}\ }\textbf {\bibinfo {volume} {99}},\
  \bibinfo {pages} {235417} (\bibinfo {year} {2019})}\BibitemShut {NoStop}%
\bibitem [{\citenamefont {Choi}\ and\ \citenamefont
  {Choi}(2019)}]{TDBG_flatband2}%
  \BibitemOpen
  \bibfield  {author} {\bibinfo {author} {\bibfnamefont {Y.~W.}\ \bibnamefont
  {Choi}}\ and\ \bibinfo {author} {\bibfnamefont {H.~J.}\ \bibnamefont
  {Choi}},\ }\href {\doibase 10.1103/PhysRevB.100.201402} {\bibfield  {journal}
  {\bibinfo  {journal} {Phys. Rev. B}\ }\textbf {\bibinfo {volume} {100}},\
  \bibinfo {pages} {201402} (\bibinfo {year} {2019})}\BibitemShut {NoStop}%
\bibitem [{\citenamefont {Koshino}(2019)}]{TDBG_topol}%
  \BibitemOpen
  \bibfield  {author} {\bibinfo {author} {\bibfnamefont {M.}~\bibnamefont
  {Koshino}},\ }\href {\doibase 10.1103/PhysRevB.99.235406} {\bibfield
  {journal} {\bibinfo  {journal} {Phys. Rev. B}\ }\textbf {\bibinfo {volume}
  {99}},\ \bibinfo {pages} {235406} (\bibinfo {year} {2019})}\BibitemShut
  {NoStop}%
\bibitem [{\citenamefont {Liu}\ \emph {et~al.}(2019)\citenamefont {Liu},
  \citenamefont {Ma}, \citenamefont {Gao},\ and\ \citenamefont
  {Dai}}]{TMG_chern}%
  \BibitemOpen
  \bibfield  {author} {\bibinfo {author} {\bibfnamefont {J.}~\bibnamefont
  {Liu}}, \bibinfo {author} {\bibfnamefont {Z.}~\bibnamefont {Ma}}, \bibinfo
  {author} {\bibfnamefont {J.}~\bibnamefont {Gao}}, \ and\ \bibinfo {author}
  {\bibfnamefont {X.}~\bibnamefont {Dai}},\ }\href {\doibase
  10.1103/PhysRevX.9.031021} {\bibfield  {journal} {\bibinfo  {journal} {Phys.
  Rev. X}\ }\textbf {\bibinfo {volume} {9}},\ \bibinfo {pages} {031021}
  (\bibinfo {year} {2019})}\BibitemShut {NoStop}%
\bibitem [{\citenamefont {Brown}\ \emph {et~al.}(2012)\citenamefont {Brown},
  \citenamefont {Hovden}, \citenamefont {Huang}, \citenamefont {Wojcik},
  \citenamefont {Muller},\ and\ \citenamefont {Park}}]{TMG_growth}%
  \BibitemOpen
  \bibfield  {author} {\bibinfo {author} {\bibfnamefont {L.}~\bibnamefont
  {Brown}}, \bibinfo {author} {\bibfnamefont {R.}~\bibnamefont {Hovden}},
  \bibinfo {author} {\bibfnamefont {P.}~\bibnamefont {Huang}}, \bibinfo
  {author} {\bibfnamefont {M.}~\bibnamefont {Wojcik}}, \bibinfo {author}
  {\bibfnamefont {D.~A.}\ \bibnamefont {Muller}}, \ and\ \bibinfo {author}
  {\bibfnamefont {J.}~\bibnamefont {Park}},\ }\href {\doibase
  10.1021/nl204547v} {\bibfield  {journal} {\bibinfo  {journal} {Nano Letters}\
  }\textbf {\bibinfo {volume} {12}},\ \bibinfo {pages} {1609} (\bibinfo {year}
  {2012})}\BibitemShut {NoStop}%
\bibitem [{\citenamefont {Wu}\ \emph {et~al.}(2016)\citenamefont {Wu},
  \citenamefont {Wang}, \citenamefont {Li}, \citenamefont {Peng},\ and\
  \citenamefont {Tan}}]{TMG_growth_carbon}%
  \BibitemOpen
  \bibfield  {author} {\bibinfo {author} {\bibfnamefont {J.-B.}\ \bibnamefont
  {Wu}}, \bibinfo {author} {\bibfnamefont {H.}~\bibnamefont {Wang}}, \bibinfo
  {author} {\bibfnamefont {X.-L.}\ \bibnamefont {Li}}, \bibinfo {author}
  {\bibfnamefont {H.}~\bibnamefont {Peng}}, \ and\ \bibinfo {author}
  {\bibfnamefont {P.-H.}\ \bibnamefont {Tan}},\ }\href
  {http://www.sciencedirect.com/science/article/pii/S0008622316307576}
  {\bibfield  {journal} {\bibinfo  {journal} {Carbon}\ }\textbf {\bibinfo
  {volume} {110}},\ \bibinfo {pages} {225 } (\bibinfo {year}
  {2016})}\BibitemShut {NoStop}%
\bibitem [{\citenamefont {Slater}\ and\ \citenamefont {Koster}(1954)}]{LCAO}%
  \BibitemOpen
  \bibfield  {author} {\bibinfo {author} {\bibfnamefont {J.~C.}\ \bibnamefont
  {Slater}}\ and\ \bibinfo {author} {\bibfnamefont {G.~F.}\ \bibnamefont
  {Koster}},\ }\href {\doibase 10.1103/PhysRev.94.1498} {\bibfield  {journal}
  {\bibinfo  {journal} {Phys. Rev.}\ }\textbf {\bibinfo {volume} {94}},\
  \bibinfo {pages} {1498} (\bibinfo {year} {1954})}\BibitemShut {NoStop}%
\bibitem [{\citenamefont {Trambly~de Laissardi\`ere}\ \emph
  {et~al.}(2012)\citenamefont {Trambly~de Laissardi\`ere}, \citenamefont
  {Mayou},\ and\ \citenamefont {Magaud}}]{TBG_TB_prove1}%
  \BibitemOpen
  \bibfield  {author} {\bibinfo {author} {\bibfnamefont {G.}~\bibnamefont
  {Trambly~de Laissardi\`ere}}, \bibinfo {author} {\bibfnamefont
  {D.}~\bibnamefont {Mayou}}, \ and\ \bibinfo {author} {\bibfnamefont
  {L.}~\bibnamefont {Magaud}},\ }\href {\doibase 10.1103/PhysRevB.86.125413}
  {\bibfield  {journal} {\bibinfo  {journal} {Phys. Rev. B}\ }\textbf {\bibinfo
  {volume} {86}},\ \bibinfo {pages} {125413} (\bibinfo {year}
  {2012})}\BibitemShut {NoStop}%
\bibitem [{\citenamefont {Huder}\ \emph {et~al.}(2018)\citenamefont {Huder},
  \citenamefont {Artaud}, \citenamefont {Le~Quang}, \citenamefont
  {de~Laissardi\`ere}, \citenamefont {Jansen}, \citenamefont {Lapertot},
  \citenamefont {Chapelier},\ and\ \citenamefont {Renard}}]{TBG_TB_prove2}%
  \BibitemOpen
  \bibfield  {author} {\bibinfo {author} {\bibfnamefont {L.}~\bibnamefont
  {Huder}}, \bibinfo {author} {\bibfnamefont {A.}~\bibnamefont {Artaud}},
  \bibinfo {author} {\bibfnamefont {T.}~\bibnamefont {Le~Quang}}, \bibinfo
  {author} {\bibfnamefont {G.~T.}\ \bibnamefont {de~Laissardi\`ere}}, \bibinfo
  {author} {\bibfnamefont {A.~G.~M.}\ \bibnamefont {Jansen}}, \bibinfo {author}
  {\bibfnamefont {G.}~\bibnamefont {Lapertot}}, \bibinfo {author}
  {\bibfnamefont {C.}~\bibnamefont {Chapelier}}, \ and\ \bibinfo {author}
  {\bibfnamefont {V.~T.}\ \bibnamefont {Renard}},\ }\href {\doibase
  10.1103/PhysRevLett.120.156405} {\bibfield  {journal} {\bibinfo  {journal}
  {Phys. Rev. Lett.}\ }\textbf {\bibinfo {volume} {120}},\ \bibinfo {pages}
  {156405} (\bibinfo {year} {2018})}\BibitemShut {NoStop}%
\bibitem [{\citenamefont {Shi}\ \emph {et~al.}(2020)\citenamefont {Shi},
  \citenamefont {Zhan}, \citenamefont {Qi}, \citenamefont {Huang},
  \citenamefont {Veen}, \citenamefont {Silva-Guill{\'e}n}, \citenamefont
  {Zhang}, \citenamefont {Li}, \citenamefont {Xie}, \citenamefont {Ji},
  \citenamefont {Katsnelson}, \citenamefont {Yuan}, \citenamefont {Qin},\ and\
  \citenamefont {Zhang}}]{TBG_TB_zhen}%
  \BibitemOpen
  \bibfield  {author} {\bibinfo {author} {\bibfnamefont {H.}~\bibnamefont
  {Shi}}, \bibinfo {author} {\bibfnamefont {Z.}~\bibnamefont {Zhan}}, \bibinfo
  {author} {\bibfnamefont {Z.}~\bibnamefont {Qi}}, \bibinfo {author}
  {\bibfnamefont {K.}~\bibnamefont {Huang}}, \bibinfo {author} {\bibfnamefont
  {E.~v.}\ \bibnamefont {Veen}}, \bibinfo {author} {\bibfnamefont {J.~{\'A}.}\
  \bibnamefont {Silva-Guill{\'e}n}}, \bibinfo {author} {\bibfnamefont
  {R.}~\bibnamefont {Zhang}}, \bibinfo {author} {\bibfnamefont
  {P.}~\bibnamefont {Li}}, \bibinfo {author} {\bibfnamefont {K.}~\bibnamefont
  {Xie}}, \bibinfo {author} {\bibfnamefont {H.}~\bibnamefont {Ji}}, \bibinfo
  {author} {\bibfnamefont {M.~I.}\ \bibnamefont {Katsnelson}}, \bibinfo
  {author} {\bibfnamefont {S.}~\bibnamefont {Yuan}}, \bibinfo {author}
  {\bibfnamefont {S.}~\bibnamefont {Qin}}, \ and\ \bibinfo {author}
  {\bibfnamefont {Z.}~\bibnamefont {Zhang}},\ }\href {\doibase
  10.1038/s41467-019-14207-w} {\bibfield  {journal} {\bibinfo  {journal}
  {Nature Communications}\ }\textbf {\bibinfo {volume} {11}},\ \bibinfo {pages}
  {371} (\bibinfo {year} {2020})}\BibitemShut {NoStop}%
\bibitem [{\citenamefont {Yuan}\ \emph {et~al.}(2010)\citenamefont {Yuan},
  \citenamefont {De~Raedt},\ and\ \citenamefont {Katsnelson}}]{TBPM}%
  \BibitemOpen
  \bibfield  {author} {\bibinfo {author} {\bibfnamefont {S.}~\bibnamefont
  {Yuan}}, \bibinfo {author} {\bibfnamefont {H.}~\bibnamefont {De~Raedt}}, \
  and\ \bibinfo {author} {\bibfnamefont {M.~I.}\ \bibnamefont {Katsnelson}},\
  }\href {\doibase 10.1103/PhysRevB.82.115448} {\bibfield  {journal} {\bibinfo
  {journal} {Phys. Rev. B}\ }\textbf {\bibinfo {volume} {82}},\ \bibinfo
  {pages} {115448} (\bibinfo {year} {2010})}\BibitemShut {NoStop}%
\bibitem [{\citenamefont {Koshino}\ and\ \citenamefont {Ando}(2008)}]{LL_AB}%
  \BibitemOpen
  \bibfield  {author} {\bibinfo {author} {\bibfnamefont {M.}~\bibnamefont
  {Koshino}}\ and\ \bibinfo {author} {\bibfnamefont {T.}~\bibnamefont {Ando}},\
  }\href {\doibase 10.1103/PhysRevB.77.115313} {\bibfield  {journal} {\bibinfo
  {journal} {Phys. Rev. B}\ }\textbf {\bibinfo {volume} {77}},\ \bibinfo
  {pages} {115313} (\bibinfo {year} {2008})}\BibitemShut {NoStop}%
\bibitem [{\citenamefont {Brenner}\ \emph {et~al.}(2002)\citenamefont
  {Brenner}, \citenamefont {Shenderova}, \citenamefont {Harrison},
  \citenamefont {Stuart}, \citenamefont {Ni},\ and\ \citenamefont
  {Sinnott}}]{reb0}%
  \BibitemOpen
  \bibfield  {author} {\bibinfo {author} {\bibfnamefont {D.~W.}\ \bibnamefont
  {Brenner}}, \bibinfo {author} {\bibfnamefont {O.~A.}\ \bibnamefont
  {Shenderova}}, \bibinfo {author} {\bibfnamefont {J.~A.}\ \bibnamefont
  {Harrison}}, \bibinfo {author} {\bibfnamefont {S.~J.}\ \bibnamefont
  {Stuart}}, \bibinfo {author} {\bibfnamefont {B.}~\bibnamefont {Ni}}, \ and\
  \bibinfo {author} {\bibfnamefont {S.~B.}\ \bibnamefont {Sinnott}},\ }\href
  {\doibase 10.1088/0953-8984/14/4/312} {\bibfield  {journal} {\bibinfo
  {journal} {Journal of Physics: Condensed Matter}\ }\textbf {\bibinfo {volume}
  {14}},\ \bibinfo {pages} {783} (\bibinfo {year} {2002})}\BibitemShut
  {NoStop}%
\bibitem [{\citenamefont {Kolmogorov}\ and\ \citenamefont
  {Crespi}(2005)}]{KC_potential}%
  \BibitemOpen
  \bibfield  {author} {\bibinfo {author} {\bibfnamefont {A.~N.}\ \bibnamefont
  {Kolmogorov}}\ and\ \bibinfo {author} {\bibfnamefont {V.~H.}\ \bibnamefont
  {Crespi}},\ }\href {\doibase 10.1103/PhysRevB.71.235415} {\bibfield
  {journal} {\bibinfo  {journal} {Phys. Rev. B}\ }\textbf {\bibinfo {volume}
  {71}},\ \bibinfo {pages} {235415} (\bibinfo {year} {2005})}\BibitemShut
  {NoStop}%
\bibitem [{\citenamefont {Plimpton}(1995)}]{lammps}%
  \BibitemOpen
  \bibfield  {author} {\bibinfo {author} {\bibfnamefont {S.}~\bibnamefont
  {Plimpton}},\ }\href {\doibase https://doi.org/10.1006/jcph.1995.1039}
  {\bibfield  {journal} {\bibinfo  {journal} {Journal of Computational
  Physics}\ }\textbf {\bibinfo {volume} {117}},\ \bibinfo {pages} {1 }
  (\bibinfo {year} {1995})}\BibitemShut {NoStop}%
\bibitem [{sm()}]{sm}%
  \BibitemOpen
  \href@noop {} {\emph {\bibinfo {title} {See Supplemental Material at link for
  SI. The validation of 15/26 approximant. SII. The interlayer decoupling in
  30{\degree} AB/G. SIII. The interlayer coupling at Q points in 30{\degree}
  AB/G.}}}\BibitemShut {Stop}%
\bibitem [{\citenamefont {Nishi}\ \emph {et~al.}(2017)\citenamefont {Nishi},
  \citenamefont {Matsushita},\ and\ \citenamefont {Oshiyama}}]{BandUnfold_SF}%
  \BibitemOpen
  \bibfield  {author} {\bibinfo {author} {\bibfnamefont {H.}~\bibnamefont
  {Nishi}}, \bibinfo {author} {\bibfnamefont {Y.-i.}\ \bibnamefont
  {Matsushita}}, \ and\ \bibinfo {author} {\bibfnamefont {A.}~\bibnamefont
  {Oshiyama}},\ }\href {\doibase 10.1103/PhysRevB.95.085420} {\bibfield
  {journal} {\bibinfo  {journal} {Phys. Rev. B}\ }\textbf {\bibinfo {volume}
  {95}},\ \bibinfo {pages} {085420} (\bibinfo {year} {2017})}\BibitemShut
  {NoStop}%
\bibitem [{\citenamefont {Medeiros}\ \emph {et~al.}(2014)\citenamefont
  {Medeiros}, \citenamefont {Stafstr\"om},\ and\ \citenamefont
  {Bj\"ork}}]{BandUnfold_EBS}%
  \BibitemOpen
  \bibfield  {author} {\bibinfo {author} {\bibfnamefont {P.~V.~C.}\
  \bibnamefont {Medeiros}}, \bibinfo {author} {\bibfnamefont {S.}~\bibnamefont
  {Stafstr\"om}}, \ and\ \bibinfo {author} {\bibfnamefont {J.}~\bibnamefont
  {Bj\"ork}},\ }\href {\doibase 10.1103/PhysRevB.89.041407} {\bibfield
  {journal} {\bibinfo  {journal} {Phys. Rev. B}\ }\textbf {\bibinfo {volume}
  {89}},\ \bibinfo {pages} {041407} (\bibinfo {year} {2014})}\BibitemShut
  {NoStop}%
\end{thebibliography}%
\end{document}